\documentclass[11pt,draftcls,onecolumn]{IEEEtran}
\usepackage{graphicx,amssymb,stfloats,color,microtype,dsfont}
\usepackage [cmex10]{amsmath}
\usepackage[noadjust]{cite}
\usepackage[hidelinks, bookmarksnumbered=true]{hyperref}
\hypersetup{pdfpagemode=UseNone, pdfstartview=FitH}

\ifCLASSOPTIONcompsoc
\usepackage[caption=false,font=normalsize,labelfont=sf,textfont=sf]{subfig}
\else
\usepackage[caption=false,font=footnotesize]{subfig}
\fi

\usepackage{mathtools}
\newlength{\mymathln}
\newcommand{\aligninside}[2]
{
  \settowidth{\mymathln}{#2}
  \mathmakebox[\mymathln]{#1}
}

\newcounter{mytempeqncnt2}

\IEEEaftertitletext{\vspace{-2\baselineskip}}

\cleardoublepage
\makeatletter
\newsavebox\myboxA
\newsavebox\myboxB
\newlength\mylenA
\newcommand*\xoverline[2][0.6]{%
    \sbox{\myboxA}{$\m@th#2$}%
    \setbox\myboxB\null
    \ht\myboxB=\ht\myboxA%
    \dp\myboxB=\dp\myboxA%
    \wd\myboxB=#1\wd\myboxA
    \sbox\myboxB{$\m@th\overline{\copy\myboxB}$}
    \setlength\mylenA{\the\wd\myboxA}
    \addtolength\mylenA{-\the\wd\myboxB}%
    \ifdim\wd\myboxB<\wd\myboxA%
       \rlap{\hskip 0.5\mylenA\usebox\myboxB}{\usebox\myboxA}%
    \else
        \hskip -0.5\mylenA\rlap{\usebox\myboxA}{\hskip 0.5\mylenA\usebox\myboxB}%
    \fi}
\makeatother

\newcommand*{\wtilde}[2][0pt]{
  \raisebox{#1}{$\widetilde{\raisebox{-#1}{$#2$}}$}%
}%

\newcommand{\colorR}{} 

\newcommand{\colorL}{}%
\newcommand{\colorH}{}
\newcommand{\colorT}{}

\let\originalleft\left
\let\originalright\right
\renewcommand{\left}{\mathopen{}\mathclose\bgroup\originalleft}
\renewcommand{\right}{\aftergroup\egroup\originalright}

\newcommand*\diff{\mathop{}\!\!\mathrm{d}}

\begin{document}

\bstctlcite{IEEEexample:BSTcontrol}

\title{Ambiguity Function of the Transmit Beamspace-Based MIMO Radar}

\author{Yongzhe Li, Sergiy A. Vorobyov, and Visa Koivunen
\thanks{Y. Li, S. A. Vorobyov and V. Koivunen are with the Department of Signal Processing and Acoustics, Aalto University, P.O. Box 13000, FI-00076 Aalto, Finland. Y. Li is also with the Department of Electronic Engineering, University of Electronic Science and Technology of China, Chengdu, 611731, China (e-mail: \href{mailto:lyzlyz888@gmail.com}{lyzlyz888@gmail.com}/\href{mailto:yongzhe.li@aalto.fi} {yongzhe.li@aalto.fi}; \href{mailto:svor@ieee.org}{svor@ieee.org}; \href{mailto:visa.koivunen@aalto.fi}{visa.koivunen@aalto.fi}).}
\thanks{Y. Li's work is supported by China Scholarship Council and by the Fundamental Research Funds for the Central Universities of China under Contract ZYGX2010YB007.}
\thanks{Some of the preliminary and relevant results on AF for MIMO radar with correlated waveforms have been presented at ICASSP, Florence, Italy, 2014.}

%
}

\maketitle

\begin{abstract}
In this paper, we derive an ambiguity function (AF) for the transmit beamspace (TB)-based multiple-input multiple-output (MIMO) radar \colorR{for the case of far-field targets and narrow-band waveforms. The effects of \colorT{transmit} coherent processing gain and waveform diversity are incorporated into the AF definition. To cover all the phase information conveyed by different factors, we introduce the equivalent transmit phase centers. The newly defined AF serves as a generalized AF form for which the phased-array \colorT{(PA)} and traditional MIMO radar AFs are important special cases. We establish relationships among the defined TB-based MIMO radar AF and the existing AF results including the Woodward's AF, the AFs defined for the traditional colocated MIMO radar, and also the PA radar AF, respectively. Moreover, we compare the TB-based MIMO radar AF with the square-summation-form AF definition and identify two limiting cases to bound its ``clear region'' in Doppler-delay domain that is free of sidelobes. Corresponding bounds for these two cases are derived, and it is shown that the bound for the worst case is inversely proportional to the number of transmitted waveforms $K$, whereas the bound \colorT{for the best case} is independent of $K$. The actual ``clear region'' of the TB-based MIMO radar AF depends on the array configuration and is in between of the worst- and best-case bounds. We propose a TB design strategy to reduce the levels of the AF sidelobes, and show in simulations that proper design of the TB matrix leads to reduction of the relative sidelobe levels of the TB-based MIMO radar AF.}
\end{abstract}


\section{Introduction}

     \colorL{The multiple-input multiple-output (MIMO) radar has become the focus of intensive research in recent years \colorH{\cite{LiMIMORadar, HaimovichMIMO, LiMIMOMag., FishlerSpatial, FishlerMIMOIdea, BlissMIMO03, BekkermanTargetDetec06, AbramovichMIMOLimit10, LiMIMOSARImaging07, KriegerMIMOSAR14}.} It has been shown that the traditional MIMO radar with colocated antenna elements enables many benefits such as \colorH{increased upper limit on the number of resolvable targets \cite{LiMIMORadar}, improved parameter identifiability and angular resolution \cite{LiMIMOMag.},} extended array aperture by virtual sensors \cite{BlissMIMO03}, \colorH{possibility of building a noncausal system for clutter mitigation \cite{AbramovichMIMOLimit10}, and capability on jammer suppression \cite{YongzheJamICASSP, YongzheRobustBeam}. Moreover,} the techniques that aim to combine the benefits of the traditional MIMO radar and the well developed phased-array (PA) radar \colorH{have emerged} in the past few years \cite{FuhrmannHybrid, VorobyovTransmit, FuhrmannTrasnBeam, StoicaProbingDesing, VorobyovPMIMO, VorobyovTBDesign, VorobyovTRBeamDesign, VbyovBeamDesignICASSP, AittomakiLowComTB07, AittomakiCovTB07}. These techniques, namely, transmit beamspace (TB) design techniques, trade off the omnidirectional transmission of mutually orthogonal waveforms to higher transmit coherent processing gain in MIMO radar. \colorH{For example,}} the work of \cite{FuhrmannHybrid} attempts to simultaneously \colorR{incorporate} the benefits of waveform diversity and \colorT{transmit} coherent processing gain by separating the transmit antenna array into several uniform subarrays, and \colorR{enabling each one} to perform as a PA. Unlike \cite{FuhrmannHybrid}, the transmit beamspace (TB)-based MIMO radar (see for example \cite{VorobyovTransmit}) focuses the energy of multiple transmitted orthogonal waveforms within a certain spatial sector where a target is likely to be located using \colorR{beamspace design techniques.} \colorR{In this radar configuration, beams that fully cover the sector-of-interest are synthesized at the transmitting end. Each beam} associated with a certain \colorR{orthogonal} waveform is implemented \colorR{via} the whole transmit array of the TB-based MIMO radar. The essence of it is to find the jointly optimal scheme \colorR{that achieves improved signal-to-noise ratio (SNR) together with increased aperture by means of TB processing techniques \cite{VorobyovTransmit, FuhrmannTrasnBeam, StoicaProbingDesing, VorobyovPMIMO, VorobyovTBDesign, VorobyovTRBeamDesign, VbyovBeamDesignICASSP, AittomakiLowComTB07, AittomakiCovTB07}. For example, it allows to achieve \colorT{transmit} coherent processing gain or desired beampattern by appropriate design of waveform correlation matrix \cite{FuhrmannTrasnBeam}, \cite{StoicaProbingDesing}.}

\colorR{Compared to the traditional MIMO radar, one verified benefit of the TB-based MIMO radar is the superior direction-of-arrival (DOA) estimation performance in a wide range of SNRs \cite{VorobyovTransmit}, \cite{VorobyovTBDesign}, \cite{VorobyovTRBeamDesign}. Based on classic MUSIC \cite{SchmidtDOA} or ESPRIT \cite{RoyESPRIT} approaches, multiple efficient algorithms that facilitate DOA estimation can be developed. Moreover, the Cram\'{e}r-Rao bound (CRB) derived for the TB-based MIMO radar in \cite{VorobyovTransmit} demonstrates that it can achieve a lower CRB with fewer waveforms than the traditional MIMO radar with full waveform diversity, and the lowest CRB can be achieved with proper TB design. This leads to emitting non-orthogonal or correlated waveforms from different transmit antenna elements. To study the performance of these actually emitted waveforms as well as the resolution performance of the TB-based MIMO radar system, it is essential to employ ambiguity function (AF) \cite{PriceBounds, SielertSelfTr, AFSLooking65, AntonioMIMOAbmF, ChenAFWaveform08, AbramovichBounds} for the performance evaluation.}

\colorR{The well-known Woodward's AF \cite{PriceBounds}, \cite{SielertSelfTr}, which characterizes the resolution property in Doppler-delay domain for narrow-band waveforms, has served as a starting point for the works on the traditional MIMO radar AF \cite{AntonioMIMOAbmF, ChenAFWaveform08, AbramovichBounds}. It has been extended to the traditional MIMO radar setup in \cite{AntonioMIMOAbmF} for the first time, and four AF simplifications corresponding to different scenarios have been derived there. Some properties of  the traditional MIMO radar AF have been studied in \cite{ChenAFWaveform08}. Another AF definition for the traditional MIMO radar which does not consider the phase information, has been introduced in \cite{AbramovichBounds}. However, with the development of TB design techniques, which allow for non-orthogonal or correlated waveforms to be emitted from each transmit antenna element, the traditional MIMO radar AFs are no longer applicable \colorT{to} the TB-based MIMO radar. This motivates us to derive the AF for the TB-based MIMO radar and investigate how it behaves. Moreover, in-depth study of the TB-based MIMO radar AF also provides insights into the clutter/interference mitigation in airborne\colorL{/spaceborne} MIMO radar system with TB design. On the other hand, it is known that the so-called ``clear region" \cite{PriceBounds}, \cite{SielertSelfTr} denotes the volume-clearance area in Doppler-delay domain which is free of sidelobes. It serves as a measure to determine how close to the ideal thumbtack-shape AF one can come. It is also of great significance for the TB-based MIMO radar AF analysis to see how large its ``clear region'' is. The work in \cite{AbramovichBounds} defines the traditional MIMO radar AF as the sum of the squared noise-free outputs after matched filtering to the waveforms. Based on this definition the ``clear region'' bound is derived. Such bound is also important to derive for the TB-based MIMO radar AF.
}

\colorR{In this paper, we derive the AF for the TB-based MIMO radar, and it serves as a generalized AF form for which the existing traditional MIMO radar AF and PA radar AF are important special cases.\footnote{Some of the preliminary and relevant results on AF for MIMO radar with correlated waveforms have been presented in \cite{YongzheAFICASSP}.} The effects of both \colorT{transmit} coherent processing gain and waveform diversity are considered when defining the new AF for the TB-based MIMO radar.  The phase information conveyed by multiple factors such as array geometry and relative motion is incorporated. Considering that it is impossible to give an exact ``clear region'' bound for the TB-based MIMO radar AF because the self-transform \cite{PriceBounds} of the TB-based MIMO radar AF can not guarantee the non-negativity in general, we identify two limiting cases to conduct the analysis.}

\colorR{The main contributions of this paper are as follows:
\vspace*{-1pt}
	\begin{itemize}
	  	\item We introduce a new AF definition for the TB-based MIMO radar for the case of far-field targets and narrow-band waveforms. Equivalent transmit phase centers are introduced in the definition as well.
   		\item We show that the TB-based MIMO radar AF is a generalization of AF for many well-known radar configurations such as the PA radar, the traditional MIMO radar (with subarrays), and the TB-based MIMO radar, by properly selecting the TB matrix and the equivalent transmit phase centers.
   		\item We establish the relationships among the defined TB-based MIMO radar AF and other existing AFs in the literature including the well-known Woodward's AF, the traditional MIMO radar AF, and the PA radar AF, respectively.
   		\item We compare the newly defined TB-based MIMO radar AF with the square-summation-form AF \cite{AbramovichBounds}, and propose a TB design strategy to reduce  the relative sidelobe levels of the TB-based MIMO radar AF.
   		\item We identify the worst and the best limiting cases for the TB-based MIMO radar AF, and derive the corresponding ``clear region'' bounds.
 	\end{itemize}}

The rest of the paper is organized as follows. In Section~\ref{Sec:SigModel}, the signal models for the traditional and the TB-based MIMO radars are briefly introduced as well \colorR{as the some preliminaries about the TB matrix design.} In Section~\ref{Sec:DefTBAMBF}, we present the definition of the TB-based MIMO radar AF \colorR{and establish} connections to the previous AF works. \colorR{A new TB design strategy that enables lower relative AF sidelobe levels is proposed in this section.} The ``clear region'' analysis for the TB-based MIMO radar AF is given in Section~\ref{Sec:AnalyTBAMBF}. In Section~\ref{Sec:Simulation}, simulation results corresponding to different types of radar AFs are provided. Polyphase-coded \cite{LevanonRadarSignal} and Gaussian sequences with single pulse are employed as the transmitted waveforms in this section. Finally, conclusions are drawn in Section \ref{Sec:Conclu}.

\section{Signal Model \colorR{and Preliminaries}}\label{Sec:SigModel}

Consider a colocated MIMO radar system with a transmit array of $M$ antenna elements and a receive array of $N$ antenna elements. \colorR{Both the transmit and receive arrays are assumed to be closely located, therefore, they share an identical spatial angle for a far-field target. In the context of the traditional MIMO radar,} the complex envelope of the waveforms emitted by the transmit antenna elements can be modeled as
\begin{equation}
s_m \colorR{\left(\tilde{\hspace*{-.2pt} t}\right)}  	=	\sqrt{\frac{E}{M}}		\phi_m \colorR{\left(\tilde{\hspace*{-.2pt} t}\right)} , \;	m=1,2,\ldots,M
\end{equation}
where $E$ is the total transmit energy \colorR{within one radar pulse,} $\colorR{\tilde{t}}$ is the continuous \colorR{fast-}time index, i.e., time within the pulse, and $\phi_m \colorR{\left(\tilde{\hspace*{-.2pt} t}\right)}$ is the $m$th orthogonal baseband waveform. Without loss of generality, we assume that the transmitted waveforms are normalized to have unit-energy, i.e.,
\begin{equation}
\int_{\colorL{T_p}} \! \vert  \phi_m \colorR{\left(\tilde{\hspace*{-.2pt} t}\right)}  \vert^2 	\diff	\colorR{\tilde{t}}   =	1, \;  m=1,2, \ldots, M
\end{equation}
where \colorL{$T_p$} is the time duration of the pulse.

Assuming that $L$ targets are present, the $N\times1$ received complex signal vector can be expressed as
\begin{equation}
\mathbf{x} \left( t,\varsigma \right) 	=\sum^{L}_{l=1} 	r_l \left( t,\varsigma\right) \mathbf{b} \left( \theta_l \right) +\mathbf{z} \left( t,\varsigma\right)
\end{equation}
where \colorR{$t$ is the continuous fast-time index for the received signal,} $\varsigma$ is the \colorR{slow-time} index, i.e., the pulse number, $\mathbf{b} (\theta_l )$ is the steering vector of the receive array associated with the $l$th target, $\mathbf{z} ( t,\varsigma ) $ is $N\times1$ zero-mean white Gaussian noise, and
\begin{equation}\label{eq:tarSig}
r_l (t,\varsigma) 	=		\sqrt{\frac {E}{M}}		\alpha_l (\varsigma) 		\colorR{D_l (\varsigma)} 	\mathbf{a}^{\colorL{\text T}} ( \theta_l )		\boldsymbol{\phi}( t )
\end{equation}
is the echo of radar return due to the $l$th target located at the spatial direction $\theta_l$. In \eqref{eq:tarSig}, $\alpha_l (\varsigma)$, $\colorR{D_l ( \varsigma )}$, $\mathbf{a} (\theta_l)$, and $\theta_l$ are respectively the complex reflection coefficient, \colorR{the phase due to Doppler,} the steering vector of transmit array, and the spatial angle all associated with the $l$th target, $\boldsymbol{\phi} (t) 	\triangleq 	[\phi_1(t),\ldots,\phi_M(t)]^{\colorL{\text T}}$ is the $M\times 1$ waveform vector, and $(\cdot)^{\colorL{\text T}}$ stands for the transpose operation. \colorL{The reflection coefficient $\alpha_l (\varsigma)$ is assumed to be constant over the whole radar coherent processign interval. \colorH{The phase term} $D_l ( \varsigma )$ is assumed to be constant for any give $t$ during the $\varsigma$th pulse, i.e., slow-moving targets are assumed.}

At the receiving end, the $N\times1$ component of the received data~\eqref{eq:tarSig} due to the $m$th waveform is extracted by employing the matched filtering technique, i.e.,
\begin{equation}\label{eq:Matsig}
\mathbf{x}_m (\varsigma ) 	\triangleq 		\int_{\colorL{T_p}} \! \mathbf{x} (t, \varsigma) 		\phi_m^\ast \left( t\right) 	\diff t, 	\;  m=1,\ldots,M
\end{equation}
where $( \cdot )^\ast$ is the conjugation operator. By stacking all the filtered components \eqref{eq:Matsig} into a column vector, we can obtain the following $MN\times1$ virtual data vector
\begin{alignat}{2}
  \mathbf{y}_\mathrm{MIMO} (\varsigma)
  & \triangleq
  \left[ \mathbf{x}_1^{\colorL{\text T}} (\varsigma), \ldots, \mathbf{x}_M^{\colorL{\text T}} (\varsigma)
  \right]^{\colorL{\text T}} \nonumber\\
  & = \sqrt{\frac{E}{M}} \sum_{l=1}^L 	\alpha_l (\varsigma) 	\colorR{D_l ( \varsigma )} \mathbf{u}_\mathrm{MIMO} (\theta_l) 	+	 \mathbf{\tilde{z}} (\varsigma)
\end{alignat}
where $\mathbf{u}_\mathrm{MIMO} 	\triangleq	\mathbf{a} (\theta) \otimes\mathbf{b} (\theta)$ is the $MN\times1$ virtual steering vector, $\mathbf{\tilde{z}} (\varsigma)$ is the $MN\times1$ noise term whose covariance is given by $\sigma_\mathbf{z}^2\mathbf{I}_{MN}$, and $\otimes$ denotes the Kronecker product.

In the TB-based MIMO radar system, \colorR{$K\,(\text{in general}, K\leq M)$ initially orthogonal waveforms are transmitted \cite{VorobyovTransmit}. For each waveform, a transmit beam that illuminates a certain area within the pre-determined spatial angular sector-of-interest $\boldsymbol{\Omega}$ is formed. The $K$ synthesized transmit beams are designed to fully cover the spatial sector $\boldsymbol{\Omega}$. Thus, in the context of the TB-based MIMO radar,} the signal radiated towards the target located at the spatial direction $\theta$ via the $k$th transmit beam can be modeled as~\cite{VorobyovTransmit}
\begin{equation}
s_k (t) 	=	\sqrt{\frac{E}{K}}	\mathbf{c}^{\colorL{\text T}}_{k} 	\mathbf{a}(\theta)	\phi_k(t), \;	k=1,\ldots,K
	\end{equation}
where $\mathbf{c}_k$ is the $k$th column vector of the $M \times K$ TB matrix $\mathbf{C}$ with $\mathbf{C}$ being defined as
\begin{equation}
\mathbf{C} \triangleq [\mathbf{c}_1,\ldots ,\mathbf{c}_K].
\end{equation}
Technically, each column of $\mathbf{C}$ that is composed of $M$ elements is \colorT{carefully} designed to form a certain transmit beam within the sector-of-interest $\boldsymbol{\Omega}$, and the $k$th orthogonal waveform is emitted through the $k$th synthesized transmit beam. Therefore, by denoting the $m$th element of $\mathbf{c}_k$ as  $c_{mk}$, the signal $\tilde{s}_m (t)$ radiated from the $m$th transmit antenna element can be expressed as
\begin{eqnarray}\label{eq:eqlTrSig}
\tilde{s}_{m}(t) 	=	\sqrt{\frac{E}{K}}	\sum^K_{k=1} 	c_{mk}		\phi_k (t), 	\; m=1,\dots,M.
\end{eqnarray}

\colorR{There are many ways of designing the TB matrix $\mathbf{C}$. For example, one way is to maximize (or keep fixed) the energy transmitted within the sector-of-interest $\boldsymbol\Omega$ while minimizing (or keeping fixed) the energy transmitted in the out-of-sector area at the same time \cite{VorobyovTransmit}. Mathematically, the constrained optimization problem for finding $\mathbf{C}$ can be expressed as
\begin{equation}\label{eq:TBAFConvex}
  \begin{alignedat}{3}
    &\underset{\mathbf{C}}{\mathrm{min}} \; \underset{i}{\mathrm{max}}
	&& \;\,
		\left\|
    	\mathbf{C}^{\colorL{\text H}}\mathbf{a}\left(\theta_i\right)-\mathbf{d}\left(\theta_i\right)
	\right\|, \;\,  &&\theta_i\in\boldsymbol{\Omega}, \, i=1,\ldots,I\\
&\aligninside{\mathrm{s.t.}}{$\underset{\mathbf{C}}{\mathrm{min}} \; \underset{i}{\mathrm{max}}$}
&&\;\, \left\|
                \mathbf{C}^{\colorL{\text H}}\mathbf{a}\left(\theta_j\right)
                \right\|
                \leq\gamma,   &&\bar{\theta}_j\in\xoverline{\boldsymbol{\Omega}}, \, j=1,\ldots,J  \\
  \end{alignedat}
\end{equation}
where $\mathbf{d}(\theta)$ is the presumed vector of size $K \times 1$ that guarantees the desired property of transmit beamforming, $\xoverline{\boldsymbol{\Omega}}$ combines a continuum of all out-of-sector directions that lie outside $\boldsymbol{\Omega}$, $\gamma$ is the parameter of the user choice that characterizes the worst acceptable level of transmit power leakage in the out-of-sector region, $I$ and $J$ are the numbers of grids of angles within and outside the sector-of-interest $\boldsymbol{\Omega}$, respectively, \colorL{$(\cdot)^{\colorL{\text H}}$ is the conjugate transpose operator,} and $\left\|\cdot\right\|$ is the Euclidean norm. The correlated waveforms $\mathbf{S}(t) \triangleq [\tilde{s}_{1}(t), \ldots, \tilde{s}_{M}(t)]$ can also be designed directly \cite{SoltanalianTBDesign14}. To achieve good Doppler tolerance of the waveforms, spectral constraints can be enforced in the designing process \cite{RoweSpecConstraint14}. In essence, both the TB matrix design and the direct correlated waveforms design can be understood as achieving an optimal (in some \colorT{pre-determined} sense) \colorT{mixing} matrix $\mathbf{R}_{\mathrm{d}}$ that can be expressed as $\mathbf{R}_{\mathrm{d}} = \mathbf{C}\mathbf{C}^{\colorL{\text H}}$ or as $ \mathbf{R}_{\mathrm{d}}  = \mathop{\mathbb{E}}\{\mathbf{S}(t)\mathbf{S}^{\colorL{\text H}}(t)\}$ with $\mathop{\mathbb{E}}\{\cdot\}$ standing for the expectation operator. In contrast to designing the \colorT{mixing} matrix $\mathbf{R}_{\mathrm{d}}$ directly \cite{FuhrmannTrasnBeam}, the TB-based approach enables us to investigate the AF of the TB-based MIMO radar.}

\section{The TB-Based MIMO Radar AF}\label{Sec:DefTBAMBF}

In this section, \colorR{we first introduce the AF of the TB-based MIMO radar, then we establish} the relationships among the so-defined AF and the previous works on AF including the well-known Woodward's AF, the traditional MIMO radar AF, \colorR{and the PA radar AF.}

\colorH{As it \colorT{has been} shown in the previous \colorT{section,} $K$ orthogonal waveforms/beams versus $M$ transmit antenna elements are employed in the TB-based MIMO radar, and generally, $K$ is much less than $M$. \colorT{Thus,} the situation when it is required to alternate between the beamspace and the element space occurs. The $K$ initial waveforms in the beamspace correspond to $M$ compound waveforms in the element space. Generally, these $M$ compound waveforms are correlated to each other, which is \colorT{achieved} by the TB matrix $\mathbf{C}$. When it comes to the introduction \colorT{of AF} for the TB-based MIMO radar, we aim \colorT{at obtaining an expression which separates} the characteristics of the $K$ initial waveforms and the effect of correlation due to the TB matrix $\mathbf{C}$. To achieve this, we start from the beamspace waveforms $\phi_k\left( t \right), \, k=1, \ldots, K$, and follow the element space signal model \eqref{eq:eqlTrSig}. \colorT{One more benefit due to this routine is that no changes need to be done at the receiver that employs a regular bank of matched filters matching to the waveforms $\phi_k\left( t \right), \, k=1, \ldots, K$.}}


\subsection{\colorR{AF Definition and Implication}}

\colorR{We consider the most common radar scenario of far-field targets and narrow-band waveforms, and assume that the TB-based MIMO radar is operating at the frequency $f_c$. For a point target located at the position $\mathbf{p}$, the received signal at the $j$th receive antenna element before demodulation to the base band can be written as
}
\colorR{
\begin{alignat}{2} \label{eq:reSig}
\tilde{r}_{j} (t,\mathbf{p})  =  \sum^{M}_{m=1} & 	\alpha_{mj}	\tilde{s}_m (t-\tau_{mj} (\mathbf{p} ) ) \nonumber\\
& \times \mathrm{exp} \{j2\pi f_c  (t-\tau_{mj} (\mathbf{p} ) )  \}  +	\tilde{z}_{j}(t)
\end{alignat}}
where $\alpha_{mj}$ \colorR{is} the complex reflection coefficient for the $(m,j)$th transmit-receive channel, $\tau_{mj}(\mathbf{p})$ is the two-way time delay \colorR{of the $(m,j)$th transmit-receive channel} due to \colorR{the target location} at $\mathbf{p}$, \colorR{$\tilde{s}_m (t-\tau_{mj} (\mathbf{p} ))$ is the time-delayed version of $\tilde{s}_m (t)$ that has been defined in \eqref{eq:eqlTrSig},}  and $\tilde{z}_j(t)$ is the noise observed by the $j$th receive antenna element.

\colorR{Let us assume that the target is moving, and its velocity and moving direction are depicted by the vector $\mathbf{v}$. For the sake of brevity, we exploit $\boldsymbol{\Theta}$ to denote the parameter of a variable in the following derivation if it is determined by both the target position $\mathbf{p}$ and the velocity vector $\mathbf{v}$. Considering the effect of target motion on Doppler in \eqref{eq:reSig} and using also \eqref{eq:eqlTrSig}, the received signal after performing demodulation to the baseband} can be expressed as 
%
%
%
%
\begin{alignat}{2}
\hat{r}_j  (t,\mathbf{\Theta}) 	=	\, &	\sqrt{\frac{E}{K}}	\sum^M_{m=1}\sum^K_{k=1}		\alpha_{mj} 	c_{mk}		\phi_k (t-\tau_{mj}(\mathbf{p}) ) 	\mathrm{exp}	\{{-j2\pi\tau_{mj} (\mathbf{p}) (f_c+f_{mj} (\mathbf{\Theta}))}\}	 \nonumber\\	
                                                &  \times \mathrm{exp} \{ {j2\pi f_{mj} (\mathbf{\Theta})t }\}+z_j(t) \label{eq:ReceiveModel}
\end{alignat}
where $f_{mj} (\mathbf{\Theta} )$ is the Doppler shift of the target due to the $(m, j)$th transmit-receive channel and $z_j(t)$ is the white Gaussian noise with power $\sigma_z^2$ observed at the $j$th receive antenna element after demodulation.

At the receiving end, a bank of matched filters \colorR{is} employed due to the fact that the received signal \colorR{is} a sum of \colorR{the reflected echoes associated with} the known transmitted waveforms. The optimal detector is a filter matched to a specific set of target parameters. \colorR{Therefore,} by matched filtering $\hat{r}_j\left(t,\mathbf{\Theta}\right)$ to each of the waveforms $\phi_k(t), \, k=1,\ldots,K$ with a specific target parameter $\mathbf{\Theta}'$, \colorR{namely,} $\phi_k(t,\mathbf{\Theta}'), \, k=1,\ldots,K$, the received signal component associated with the $i$th transmitted waveform can be obtained as
%
%
%
%
%
\begin{alignat}{3}
  \bar{r}_{ji}  (\mathbf{\Theta},\mathbf{\Theta}')    = \;	&  \int \! {	\hat{r}_j (t,\mathbf{\Theta})  	\phi_{i}^\ast (t,\mathbf{\Theta}')  \diff t	} \nonumber \\
   = \; & 	\sqrt{\frac{E}{K}} \sum^M_{m=1}\sum^K_{k=1} 	\alpha_{mj} \int \! {	c_{mk}\phi_k(t-\tau_{mj} (\mathbf{p} ) )  \phi_{i}^\ast (t-\tau_{q(i) j} (\mathbf{p'} ) )}
  \mathrm{exp}  \left\{ -j2\pi\tau_{mj} (\mathbf{p} ) \right.  \nonumber\\
&  \times
\left. (f_c+f_{mj}(\mathbf{\Theta} ))  \right\}
\mathrm{exp} \{ {j2\pi\tau_{q (i) j} (\mathbf{p'}) (f_c+f_{q(i) j} (\mathbf{\Theta'} ))} \} \nonumber \\
& \times \mathrm{exp} \{ {j2\pi (f_{mj} (\mathbf{\Theta})-f_{q(i) j} (\mathbf{\Theta'}) )t} \}  \diff t 	+		\bar{z}_{ji} (t)	\nonumber\\
\triangleq \; & \bar{r}_{ji}'\left(\mathbf{\Theta},\mathbf{\Theta'}\right)+\bar{z}_{ji}\left(t\right)  	\label{eq:iReceiveModel}
\end{alignat}
where $q\left(i\right)$ is the equivalent \colorR{transmit} phase center
 for the $i$th transmitted waveform and $\bar{z}_{ji}(t)$ is the noise after matched filtering.

Let us define the AF as the square of coherent summation of all the noise-free matched filtering output pairs $\left(j,i\right)$, $j=1,\ldots,N$ and $i=1,\ldots,K.$ Thus, the AF of the TB-based MIMO radar can be mathematically expressed as
%
%
%
%
%
%
\begin{alignat}{2}
  \chi  (\mathbf{\Theta},\mathbf{\Theta'})  \triangleq \;   &  \left\vert \sum_{j=1}^{N}\sum_{i=1}^K    \bar{r}_{ji}' ( \mathbf{\Theta,\Theta'} )\right\vert^2 	\nonumber \\	
  = \; &	
  \left\vert \vphantom{\sum_{j=1}^N} \sqrt{\frac{E}{K}} \sum_{j=1}^N\sum_{i=1}^K\sum_{m=1}^M\sum_{k=1}^K		\alpha_{mj}	\int \! { c_{mk} 	\phi_k (t-\tau_{mj} (\mathbf{p} ) )} {\phi_{i}^\ast (t-\tau_{q(i) j} (\mathbf{p'} ) )}  \right. \! \nonumber \\
&\times  \mathrm{exp} \{ {-j2\pi\tau_{mj} (\mathbf{p}) (f_c+f_{mj}( \mathbf{\Theta} ))} \}
\mathrm{exp} \{ {j2\pi\tau_{q(i) j} (\mathbf{p'}) (f_c+f_{q(i) j} (\mathbf{\Theta'})) } \} \nonumber\\
& \times \mathrm{exp} \{ {j2\pi (f_{mj} (\mathbf{\Theta})-f_{q(i) j} (\mathbf{\Theta'} ) ) t } \}  \diff t 	\!\left.\vphantom{\sum_{j=1}^N}\right\vert^2. \label{eq:pmAmbF}
\end{alignat}
\colorR{Introducing an $M \times K$ matrix $\mathbf{R}$ whose $\left(m,i\right)$th element is defined as
}
%
%
%
%
%
%
\begin{alignat}{2}
[\mathbf{R}]_{mi}  (\mathbf{\Theta}, \mathbf{\Theta'}, \mathbf{C}, j)
\triangleq \; & \sqrt{\frac{E}{K}}\sum_{k=1}^Kc_{mk} \int \!  \phi_k (t-\tau_{mj} (\mathbf{p} ) ) \phi_{i}^\ast (t-\tau_{q (i) j} (\mathbf{p'} ) ) \nonumber \\
& \times \mathrm{exp} \{ {j2\pi (f_{mj} (\mathbf{\Theta} )-f_{q (i) j} (\mathbf{\Theta'} ) )t} \} \diff t \label{eq:RNotation}
\end{alignat}
\colorR{the TB-based MIMO radar AF \eqref{eq:pmAmbF} can be simplified as
}
%
%
%
%
%
%
%
%
\begin{alignat}{2}
  \chi  (\mathbf{\Theta}, \mathbf{\Theta'})
 = \; &
\left\vert
\sum_{j=1}^N\sum_{i=1}^K\sum_{m=1}^M	\alpha_{mj}	[\mathbf{R}]_{mi} (\mathbf{\Theta}, \mathbf{\Theta'}, \mathbf{C}, j) \mathrm{exp} \{ {-j2\pi\tau_{mj} (\mathbf{p} ) (f_c+f_{mj} (\mathbf{\Theta} ) )}\} \right. \nonumber \\
& \times \mathrm{exp} \{ {j2\pi\tau_{q (i) j} (\mathbf{p'} ) (f_c+f_{q (i) j} (\mathbf{\Theta'} ) )} \}
\! \left. \vphantom{\sum_{j=1}^N\sum_{i=1}^K\sum_{m=1}^M} \right\vert^2.  \label{eq:ambFuncFnl}
\end{alignat}

\colorR{The TB-based MIMO radar AF \eqref{eq:ambFuncFnl} is composed of square of summation terms,} and each summation term contains two more components in addition to \colorR{the complex reflection coefficient part.} One is the \colorR{match-filtered} component denoted by the matrix $\mathbf{R}$ \colorR{that has been expressed by \eqref{eq:RNotation}}, which stands for the \colorR{effect of waveform properties, i.e., the auto- and cross-correlations of the transmitted waveforms, and their Doppler tolerance.} The other component is composed of the last two exponential terms in \eqref{eq:ambFuncFnl}, \colorR{and it stands for the phase shift information due to the relative target position and motion with respect to the transmit and receive arrays.} \colorR{The TB-based MIMO radar AF \eqref{eq:ambFuncFnl}} can also be understood as follows. The $m$th transmit antenna element emits a compound signal that contains all the $K$ orthogonal waveforms, and these waveforms are windowed by the elements of the $m$th row in the TB matrix $\mathbf{C}$. Consequently, the matrix $\mathbf{R}$ should be of size $M\times K$, meaning that the TB matrix $\mathbf{C}$ has been employed to transform the original $K\times K$ \colorR{matrix of  waveform properties} to $\mathbf{R}$. \colorR{This presents the most significant difference that distinguishes the TB-based MIMO radar AF from the traditional MMO radar AF. Therefore, the AF defined in \cite{AntonioMIMOAbmF} is not applicable to the TB-based MIMO radar.}

The main objective of \colorR{incorporating} phase shift information in \eqref{eq:ambFuncFnl} is for taking into account the property of coherent processing introduced by \colorR{the colocated array geometry and the specific radar configuration. Therefore, if the $i$th equivalent transmit phase center is selected to be the position of the $i$th transmit antenna element, it matches the way of processing in the traditional MIMO radar. If the position of the first (or the reference) transmit antenna element is selected, then it matches the case in the PA radar. The equivalent transmit phase centers of the TB-based MIMO radar depend on the exact form of the TB matrix $\mathbf{C}$.} By properly designing the matrix $\mathbf{C}$ and the equivalent transmit phase centers, the AF \eqref{eq:ambFuncFnl} can serve as the AF of the PA, the traditional MIMO, and the TB-based MIMO radars. Hence, it \colorR{can be viewed as} a generalized \colorR{AF form} for the \colorR{currently} existing radar configurations.

%
%
%

\subsection{Relationships With Other AFs}

\colorR{The standard assumption of far-field targets and narrow-band waveforms is used in this paper. The antenna elements of the transmit and receive arrays have locations $\{\mathbf{q}_{\mathrm{T},1},\ldots,\mathbf{q}_{\mathrm{T},M}\}$ and $\{\mathbf{q}_{\mathrm{R},1},\ldots, \mathbf{q}_{\mathrm{R},N}\}$ in three-dimensional Cartesian coordinate system, respectively. The equivalent transmit phase centers are assumed to have locations} $\{\mathbf{q}_{\mathrm{TE},1},\ldots, \mathbf{q}_{\mathrm{TE},K}\}$. Here $\mathbf{q}_{\mathrm{T},i}, \, i=1,\ldots,M;$ $\mathbf{q}_{\mathrm{R},i}, \, i=1,\ldots,N;$ and $\mathbf{q}_{\mathrm{TE},i}, \, i=1,\ldots,K$ are all $1\times 3$ vectors. \colorR{In addition, we} let $\mathbf{u}\left(\mathbf{\Theta}\right)$ be a unit-norm \colorR{direction} vector pointing from the transmit/receive array to the target \colorR{identified by} the parameter $\mathbf{\Theta}$.

We can neglect the effect of target reflection coefficients for different transmit-receive channels, i.e., \colorR{assume} that all $\alpha_{mj}$ are equal to one.  This assumption is valid because \colorR{the contributions of transmit-receive channels to the TB-based MIMO radar AF are constant at any given time $t$ under the standard case of far-field targets and narrow-band waveforms.} The effect of $\alpha_{mj}$ on the TB-based MIMO radar AF is still constant even when considering multiple pulses and inter-pulse varying target reflection coefficients if \colorR{wide pulse} is employed and no range foldering \cite{ChenAFWaveform08} occurs. Then the AF \eqref{eq:ambFuncFnl} can be simplified as
\begin{equation}\label{eq:ambFuncFnlSimp}
\chi  \left(\mathbf{\Theta},\mathbf{\Theta'}\right)
=
\left|
	\mathbf{a}_\mathrm{R}^{\colorL{\text H}} \left(\mathbf{\Theta}\right) \mathbf{a}_\mathrm{R} \left(\mathbf{\Theta}'\right)
\right|^2
\left|
	\mathbf{a}_\mathrm{T}^{\colorL{\text H}} \left(\mathbf{\Theta}\right) \xoverline{\mathbf{R}} \mathbf{a}_\mathrm{TE} 	\left(\mathbf{\Theta'}\right)
\right|^2
\end{equation}
where the $\left(m,i\right)$th element of the $M\times K$ matrix $\xoverline{\mathbf{R}}$ is expressed as
\begin{alignat}{2}
\left[\xoverline{\mathbf{R}}\right] &  _{mi}  \left(\Delta\tau,\Delta f_d,\mathbf{C}\right)\label{eq:RMatrix}\\
& = \sqrt{\frac{E}{K}}\sum_{k=1}^Kc_{mk} \int \! {\phi_k\left(t\right)\phi_{i}^\ast\left(t-\Delta\tau\right)} \mathrm{exp}\left\{{j2\pi\Delta f_dt}\right\} \diff t. \nonumber
\end{alignat}
Here also $\Delta\tau \triangleq \tau(\mathbf{p})-\tau(\mathbf{p}')$, $\Delta f_d \triangleq f(\mathbf{\Theta})-f(\mathbf{\Theta}')$, and
\begin{alignat}{3}
& \mathbf{a}_\mathrm{T} \left( \mathbf{\Theta} \right)  && \triangleq  \left[ \mathrm{exp}\{ \tilde{\mathbf{u}}^{\colorL{\text T}} \left(\mathbf{\Theta}\right) \mathbf{q}_{\mathrm{T},1}  \},   \ldots,  \mathrm{exp}\{ {\tilde{\mathbf{u}}^{\colorL{\text T}} \left(\mathbf{\Theta}\right) \mathbf{q}_{\mathrm{T},M} } \} \right]^{\colorL{\text T}}  \label{eq:TrSteering}  \\ 
& \mathbf{a}_\mathrm{R} \left(\mathbf{\Theta}\right)
 && \triangleq
\left[
	\mathrm{exp}{ \{ \tilde{\mathbf{u}}^{\colorL{\text T}} 	\left(\mathbf{\Theta}\right) \mathbf{q}_{\mathrm{R},1} \} },
  	\ldots ,
  \mathrm{exp}{ \{ \tilde{\mathbf{u}}^{\colorL{\text T}}  \left(\mathbf{\Theta}\right) \mathbf{q}_{\mathrm{R},N} \} }
\right]^{\colorL{\text T}}  \label{eq:ReSteering} \\  
& \mathbf{a}_\mathrm{TE} \left(\mathbf{\Theta}\right) && \triangleq \left[ \mathrm{exp}\{ {\tilde{\mathbf{u}}^{\colorL{\text T}} \left(\mathbf{\Theta}\right) \mathbf{q}_{\mathrm{TE},1} } \},  \ldots ,  \mathrm{exp}\left\{\tilde{\mathbf{u}}^{\colorL{\text T}} \left(\mathbf{\Theta}\right) \mathbf{q}_{\mathrm{TE},K} \right\} \right]^{\colorL{\text T}}  \label{eq:EqTrSteering} 
\end{alignat}
are the $M\times 1$ transmit steering vector, the $N\times 1$ receive steering vector, and the $K\times 1$ equivalent transmit steering vector, respectively, with $\tilde{\mathbf{u}} (\mathbf{\Theta}) \triangleq {j2 \pi f' (\mathbf{\Theta})\cdot \mathbf{u} (\mathbf{\Theta}) }/{c}$ and $f' (\mathbf{\Theta}) \triangleq f_c + f (\mathbf{\Theta})$. The dependence of $\xoverline{\mathbf{R}}$ \colorT{on} $\Delta\tau,\Delta f_d$, and $\mathbf{C}$ is not shown in~\eqref{eq:ambFuncFnlSimp} for brevity, and the subscript indices for $\tau$ and $f$ are omitted \colorR{since we consider the case of far-field target and narrow-band waveforms.}

\colorR{It is known that} the Woodward's AF for a single waveform $u\left(t\right)$ can be expressed as
\begin{equation}\label{eq:ambfFuncWoodward}
{\xoverline{\chi}}\left(\tau,f_d\right) = \int \! {u\left(t\right)u^\ast\left(t-\tau\right)\mathrm{exp}\left\{{j2\pi f_dt}\right\} \diff t}.
\end{equation}
Based on \colorR{this expression,} we can define the $K\times K$ matrix $\boldsymbol{\xoverline{\chi}}\left(\tau,f_d\right)$ as the AF matrix of the $K$ orthogonal waveforms for the TB-based MIMO radar. \colorR{The $(j, k)$th element of $\boldsymbol{\xoverline{\chi}}\left(\tau,f_d\right)$ is given by}
\begin{equation}\label{eq:WoodwardAFMatrix}
\left[\boldsymbol{\xoverline{\chi}}\right]_{jk}\left(\tau,f_d\right) = \int \! {\phi_j\left(t\right)\phi_k^\ast\left(t-\tau\right)\mathrm{exp}\left\{{j2\pi f_dt}\right\} \diff t}.
\end{equation}
Using \eqref{eq:RMatrix} and \eqref{eq:WoodwardAFMatrix}, the AF \eqref{eq:ambFuncFnlSimp} can be expressed as
\begin{align}
  \chi  \left(\mathbf{\Theta},\mathbf{\Theta'}\right)
  = \; &\frac{E}{K}
  \left|
  		\mathbf{a}_\mathrm{R}^{\colorL{\text H}}\left(\mathbf{\Theta}\right)\mathbf{a}_\mathrm{R}\left(\mathbf{\Theta}'
  \right)\right|^2 \label{eq:ambfFuncSeprWood} \\
  &\times \left|\mathbf{a}_\mathrm{T}^{\colorL{\text H}}\left(\mathbf{\Theta}\right)\mathbf{C}\boldsymbol{\xoverline{\chi}} \left(\Delta\tau,\Delta f_d\right)\mathbf{a}_\mathrm{TE}\left(\mathbf{\Theta'}\right) \vphantom{\mathbf{a}_R^{\colorL{\text H}}}\right|^2	\nonumber
\end{align}
where $\boldsymbol{\xoverline{\chi}}\left(\Delta\tau,\Delta f_d\right)$ is the $K\times K$ matrix whose elements are \colorR{obtained from \eqref{eq:WoodwardAFMatrix} by changing the parameters $\tau$ and $f_d$ into $\Delta\tau$ and $\Delta f_d$, respectively.} \colorR{Realizing} that $\Delta\tau$ and $\Delta f_d$ depend on $\mathbf{\Theta}$ and $\mathbf{\Theta'}$, we employ these two parameters to denote the TB-based MIMO radar AF. \colorR{In the following, we show how the derived AF is a generalization of the widely used AF results for different radar configurations.}

Equation \eqref{eq:ambfFuncSeprWood} establishes the connection between the TB-based MIMO radar AF and the well known Woodward's AF. The TB matrix $\mathbf{C}$ transforms the original transmit steering vector of length $M$ into a new one of length $K$. Both the transformed and the equivalent transmit steering vectors are acting on the \colorR{Woodward} AF matrix \colorT{of the $K$ waveforms,} representing both the \colorT{transmit coherent} processing gain and the waveform diversity. Equivalently, we can say that each AF is windowed by the product of a \colorT{transmit} coherent processing gain and an equivalent transmit phase term. To be precise, for the $j$th and $k$th waveforms, the \colorR{quantity} $[\boldsymbol{\xoverline{\chi}} (\Delta\tau,\Delta f_d)]_{jk}, \, j,k\in\{1,\ldots,K\}$ is windowed by the product of the $j$th \colorT{transmit} coherent processing gain, namely, $\Upsilon_j \triangleq \mathbf{a}_{\mathrm{T}}^{\colorL{\text H}}(\boldsymbol\Theta)\mathbf{c}_j$ and the $k$th equivalent transmit phase term which is denoted by the $k$th element of $\mathbf{a}_{\mathrm{TE}}(\boldsymbol\Theta')$.

\colorR{Equation \eqref{eq:ambfFuncSeprWood} establishes the connection between the TB-based MIMO radar AF and the traditional MIMO radar AF. If} the number of transmitted waveforms $K$ is increased to $M$, $\mathbf{C}$ is simply the $M \times M$ identity matrix $\mathbf{I}_M$, and the equivalent transmit phase centers are selected to be the positions of the $M$ \colorR{individual} transmit antenna elements, then the TB-based MIMO radar AF \eqref{eq:ambfFuncSeprWood} becomes the following form
\begin{alignat}{2}
\chi   _\mathrm{MIMO}\left(\mathbf{\Theta},\mathbf{\Theta'}\right) =  \; & \frac{E}{M}  \left\vert\mathbf{a}_\mathrm{R}^{\colorL{\text H}}\left(\mathbf{\Theta}\right)\mathbf{a}_\mathrm{R} \left(\mathbf{\Theta}'\right)\right\vert^2   &\label{eq:ambfFuncTraMIMO}\\
&\times \left\vert\mathbf{a}_\mathrm{T}^{\colorL{\text H}}\left(\mathbf{\Theta}\right)\boldsymbol{\xoverline{\chi}}\left(\Delta\tau,\Delta f_d\right)\mathbf{a}_\mathrm{T}\left(\mathbf{\Theta'}\right)\vphantom{\mathbf{a}_R^{\colorL{\text H}}}\right\vert^2 \quad \nonumber
\end{alignat}
\colorR{which} denotes the traditional MIMO radar AF and has \colorR{exactly} the same form as the AF definition in \cite{AntonioMIMOAbmF} except for the magnitude term. This term represents the general expression of the transmit power allocation for the traditional MIMO radar. Therefore, if $E$ is selected to be equal to $M$, the expression \eqref{eq:ambfFuncTraMIMO} and the definition of AF in \cite{AntonioMIMOAbmF} have identical expressions. Furthermore, the TB-based MIMO radar AF  \eqref{eq:ambfFuncSeprWood} also shows compatibility with the traditional MIMO radar with $K$ uniform subarrays \cite{FuhrmannHybrid}, if $\mathbf{C}$ is properly designed to be a block diagonal TB matrix whose block diagonal elements are associated with the subarrays. The equivalent phase centers in this case are selected as the centers of subarrays.

\colorR{Equation \eqref{eq:ambfFuncSeprWood} also establishes the connection between the TB-based MIMO radar AF and the PA radar AF. If the number of transmitted waveforms $K$ is decreased to 1, $\mathbf{C}$ becomes just a beamforming weight vector $\mathbf{w}$, and the equivalent transmit phase center is selected to be the first (or the reference) transmit antenna. Then the TB-based MIMO radar AF takes the following form 
\begin{equation} \label{eq:ambfFuncPA}
\chi   _\mathrm{PA}\left(\mathbf{\Theta},\mathbf{\Theta'}\right) =   E  \left|\mathbf{a}_\mathrm{R}^{\colorL{\text H}}\left(\mathbf{\Theta}\right)\mathbf{a}_\mathrm{R} \left(\mathbf{\Theta}'\right)\right|^2   \left|\mathbf{a}_\mathrm{T}^{\colorL{\text H}}\left(\mathbf{\Theta}\right)  \mathbf{w} \xoverline{\chi}\left(\Delta\tau,\Delta f_d\right)  \right|^2
\end{equation}
where \colorT{the Woodward's AF $\xoverline{\chi} (\Delta\tau,\Delta f_d)$ for the only transmitted waveform in PA radar is obtained from \eqref{eq:ambfFuncWoodward}} by changing the parameters $\tau$ and $f_d$ into $\Delta\tau$ and $\Delta f_d$, respectively. Considering that the magnitude of the equivalent transmit phase center in the PA mode is constant, it can be neglected when deriving \eqref{eq:ambfFuncPA}. Consequently, the TB-based MIMO radar AF defined in this paper serves as a universal AF definition for the traditional MIMO radar (with subarrays) and the PA radar. Moreover, this generalized AF definition \colorR{can be} expressed using the Woodward's AF matrix which links it to the Woodward's AF.}

\colorL{
It is also worth noticing the difference between the TB-based MIMO radar AF and the traditional MIMO radar AF in \cite{AbramovichBounds} which defines it as the sum of squared noise-free match-filtered outputs. The TB-based MIMO radar AF \eqref{eq:ambfFuncSeprWood} incorporates phase shift information introduced by the array geometry and the relative motion, and furthermore exploits the square of summation of all the  auto- and cross-AF terms of the $K$ waveforms as the TB-based MIMO radar AF metric. This operation enables it to obtain lower relative sidelobe levels in the Doppler-delay domain than that of the AF in \cite{AbramovichBounds}. The reason lies in the mathematical expression itself and the waveform orthogonality.  In a word, the existing AF definitions in \cite{AntonioMIMOAbmF} and \cite{AbramovichBounds} for the traditional MIMO radar and the AF defined here for the TB-based MIMO radar all adopt the matched-filtering-based definition, and they are developed on the basis of the Woodward's AF. The former two AFs \colorH{deal} with orthogonal waveforms, while the TB-based MIMO radar AF extends it \colorH{to the situation} of waveform correlations.}

\subsection{\colorR{New TB Design}}

\colorR{The existing TB strategies are designed based on zero-Doppler and zero-delay AF cut, i.e., only spatial information is incorporated in the designs.
Therefore, we can also control the relative sidelobe levels of the TB-based MIMO radar AF by enforcing additional constraints on different Doppler and delay bins during the design process of the TB matrix $\mathbf{C}$. For example, if the relative sidelobes of the TB-based MIMO radar AF within certain Doppler and delay sectors-of-interest $\boldsymbol{\mathfrak{F}}$ and $\boldsymbol{\mathfrak{D}}$ are desired to be kept below a certain level, the TB strategy \eqref{eq:TBAFConvex} can be redesigned by solving the following optimization problem
\begin{subequations}\label{eq:TBAFConvexNew}
  \begin{alignat}{3}
    &\underset{\mathbf{C}}{\mathrm{min}} \; \underset{i}{\mathrm{max}}
	&& \;
		\left\|
    	\mathbf{C}^{\colorL{\text H}}\mathbf{a}_{\mathrm{T}}\left(\boldsymbol\theta_i\right) \odot \mathbf{a}_{\mathrm{TE}} \left(\boldsymbol\theta_i\right)   - \mathbf{d}\left(\boldsymbol\theta_i\right)
	\right\|\\
&&& \; \hphantom{\qquad\qquad\qquad\quad\,} \, \boldsymbol\theta_i\in\boldsymbol{\Omega}, \, i=1,\ldots,I \nonumber \\
&\aligninside{\mathrm{s.t.}}{$\underset{\mathbf{C}}{\mathrm{min}} \; \underset{i}{\mathrm{max}}$}
&&\; \left\|
                \mathbf{C}^{\colorL{\text H}}\mathbf{a}_{\mathrm{T}}\left(\boldsymbol\theta_j\right) \odot \mathbf{a}_{\mathrm{TE}} \left(\boldsymbol\theta_j\right)
                \right\|
                \leq\gamma\\
&&& \; \hphantom{\qquad\qquad\qquad\quad\,} \, \bar{\boldsymbol\theta}_j\in\xoverline{\boldsymbol{\Omega}}, \, j=1,\ldots,J \nonumber \\
&&& \; \left|\mathbf{a}_\mathrm{T}^{\colorL{\text H}}\left(\boldsymbol\vartheta_0, f_d^0 \right) \mathbf{C} \right.   \nonumber \\
&&& \;  \times  \boldsymbol{\xoverline{\chi}} \big( \left(\Delta\tau\right)_p, \left(\Delta f_d\right)_q \big) \mathbf{a}_\mathrm{TE}\big(\boldsymbol\vartheta_{\tilde{i}}, \left(f_d\right)_q \big) \! \left.\vphantom{\mathbf{a}_\mathrm{T}^{\colorL{\text H}}\left(\boldsymbol\vartheta_0, f_d^0 \right) \mathbf{C}} \right|  \leq \delta \\
&&& \; \hphantom{\qquad\qquad\qquad\quad\,} \, \left(\Delta\tau\right)_p \in \boldsymbol{\mathfrak{D}}, \, p = 1, \dots, P \nonumber \\
&&& \; \hphantom{\qquad\qquad\qquad\quad\,} \, \left(\Delta f_d\right)_q \in \boldsymbol{\mathfrak{F}}, \, q = 1, \dots, Q \nonumber \\
&&& \; \hphantom{\qquad\qquad\qquad\quad\,} \, \boldsymbol\vartheta_{\tilde{i}} \in \wtilde{\boldsymbol\Omega}, \, \tilde{i} = 1, \ldots, \tilde{I} \nonumber \\
&&& \; \mathbf{a}_\mathrm{T}^{\colorL{\text H}}\left(\boldsymbol\vartheta_0,  f_d^0 \right) \mathbf{C} \mathbf{a}_\mathrm{TE}\left(\boldsymbol\vartheta_0,  f_d^0  \right) = K
  \end{alignat}
\end{subequations}
where $\boldsymbol\vartheta_0$ and $ f_d^0 $ are respectively the spatial angular vector and the Doppler frequency of the target, $\wtilde{\boldsymbol\Omega}$ combines the spatial region of interest where the AF sidelobes need to be suppressed using $\tilde{I}$ grids of spatial directions $\{\boldsymbol\vartheta_{\tilde{i}} \in \wtilde{\boldsymbol\Omega}, \, \tilde{i} = 1, \ldots, \tilde{I}\}$, $\{(\Delta\tau)_p \in \boldsymbol{\mathfrak{D}}, \, p=1, \ldots, P\}$ and $\{(\Delta f_d)_q \in \boldsymbol{\mathfrak{F}}, \, q=1, \ldots, $ $Q\}$ are grids of delay and Doppler used to approximate the sectors-of-interest $\boldsymbol{\mathfrak{D}}$ and $\boldsymbol{\mathfrak{F}}$ by finite numbers of $P$ and $Q$ delay and Doppler bins, respectively, $(f_d)_q\triangleq (\Delta f_d)_q + f_d^0 $, and $\delta$ is the parameter of user choice that characterizes the sidelobe levels of the AF in the intersection of $\boldsymbol{\mathfrak{D}}$, $\boldsymbol{\mathfrak{F}}$, and $\wtilde{\boldsymbol\Omega}$.
}
\colorR{It is worth noting that for a certain set of designed waveforms and a fixed group of parameters $((\Delta\tau)_p,(\Delta f_d)_q), \, p \in \{1,\ldots,P\}$ and $  q \in \{1,\ldots,Q\}$, the matrix $\boldsymbol{\xoverline{\chi}} ((\Delta\tau)_p,(\Delta f_d)_q )$ in \eqref{eq:TBAFConvexNew} can be easily known from \eqref{eq:WoodwardAFMatrix}.} This motivates us to further explore the ``clear region'' bound of the TB-based MIMO radar AF which is studied in the next section.

\section{``Clear Region'' Analysis of the TB-Based MIMO Radar AF}\label{Sec:AnalyTBAMBF}

The Siebert's self-transform property~\cite{SielertSelfTr} \colorR{expressed by the following equality}
\begin{alignat}{2}
&\left| \xoverline{\chi}\left(\sigma,\nu\right)\right|^2 \label{eq:SiebertSelfTr}\\
& \qquad\quad =  \iint_{-\infty}^{\infty} \!  {
\left| \xoverline{\chi}\left(\tau,f_d\right)\right|^2} {\mathrm{exp}\left\{{-j2\pi\nu\tau+j2\pi  f_d\sigma}\right\}     \diff \tau \diff  f_d} \nonumber
\end{alignat}
holds for the Woodward's AF, \colorR{and it is} required that the transform \eqref{eq:SiebertSelfTr} be non-negative when conducting the ``clear region'' analysis \colorR{\cite{PriceBounds}.} Here \colorR{$\xoverline{\chi} \left(\sigma, \nu\right)$ is the new Woodward's AF generated from \eqref{eq:ambfFuncWoodward} by replacing the parameters $\tau$ and $f_d$ with $\sigma$ and $\nu$, respectively.} \colorR{In the context of the TB-based MIMO radar,} let $f(\sigma,\nu)$ denote the self-transform of \colorR{its AF $\chi (\boldsymbol{\Theta},\boldsymbol{\Theta'} )$,} i.e.,
\begin{alignat}{2}
&  f  \left(\sigma,\nu\right) \label{eq:FSelfTransform}\\
&  =  \iint_{-\infty}^{\infty}  \!  {\chi\left(\mathbf{\Theta},\mathbf{\Theta'}\right)}
{\mathrm{exp}\left\{{-j2\pi\nu\Delta\tau+j2\pi \Delta f_d\sigma}\right\}     d \Delta\tau d \Delta f_d}. \nonumber
\end{alignat}
\colorR{Normally, the TB-based MIMO radar} AF \eqref{eq:ambfFuncSeprWood} has negative terms in its expansion. \colorR{Therefore, the transform} \eqref{eq:FSelfTransform} contains negative terms. \colorR{Realizing this fact, it becomes clear that in general it is not guaranteed that $f(\sigma,\nu)$ is non-negative.} However, \colorR{it is needed in order to derive the ``clear region'' bound of the TB-based MIMO radar AF. Hence,} to see how large the maximum \colorR{achievable} ``clear region'' of the TB-based MIMO radar AF is, we identify two limiting cases which both enable $f(\sigma,\nu)$ to be non-negative. In the first case, we only consider the squared AF terms in the expansion of \eqref{eq:ambfFuncSeprWood}. It is later shown that this case achieves the smallest ``clear region'' and has high relative \colorR{sidelobe levels. Thus,} it can be considered as the \colorT{worst case} for the ``clear region'' \colorR{bound} of the TB-based MIMO radar AF defined in this paper. In the second case, we assume that all the cross-AFs of the $K$ waveforms are zero, i.e., \colorT{we ignore} the effects of the components \colorR{in the AF expansion of \eqref{eq:ambfFuncSeprWood}} that are associated with the sidelobes \colorR{resulting} from different pairs of waveforms. \colorR{This case represents the best situation for the ``clear region" bound of the TB-based MIMO radar AF. However,} it can never be achieved \colorR{because in general more than one waveforms is transmitted in the TB-based MIMO radar system.} \colorR{The actual maximum achievable ``clear region'' bound of the TB-based MIMO radar AF is in between that of these two cases,} and it depends on the level of the non-squared terms of the AF expansion which are windowed by the \colorT{transmit} coherent processing gains and the equivalent transmit phase terms.

In the remaining part of this section, we analyze the worst- and best-case ``clear region''. We first \colorR{derive the bounds for these two cases,} then we conduct the analysis based on these two bounds. \colorR{The} superscripts $(\cdot)^\text{I}$ and $(\cdot)^\text{II}$ \colorR{are used for denoting the quantities with respect to} the worst- and best-cases, respectively.

\subsection{\colorR{Worst-Case Bound}}

\colorR{In the worst-case,} in order to find the maximum \colorR{achievable sidelobe-free area in Doppler-delay domain,} we specify the following \colorR{relaxed} conditions on the auto- and cross-AFs
\begin{equation} \label{eq:AFCondtions}
	\begin{cases*}
  	\!	\iint_A { |\left[\boldsymbol{\xoverline{\chi}}\right]_{jj}\left(\tau,f_d\right) |^2  \diff \tau \diff f_d}	\simeq  \! \underset {(0,0)} {\iint}  |\left[\boldsymbol{\xoverline{\chi}}\right]_{jj}\left(\tau, f_d\right) |^2   \diff \tau \diff f_d 	\triangleq V_{j} \\
  	\!	\iint_A { |\left[\boldsymbol{\xoverline{\chi}}\right]_{jk,j\neq k} \left(\tau,f_d\right)  |^2  \diff \tau \diff f_d} \simeq 0
	\end{cases*}
\end{equation}
where $A$ denotes \colorR{the convex and centrosymmetric region of integration in  the Doppler-delay plane. Here, we define} the volume of the TB-based MIMO radar AF over the \colorR{integral} region $A$ as
\begin{equation}\label{eq:volumnDef}
V_\mathrm{TB}\left(A\right) \triangleq \iint_A{\chi\left(\mathbf{\Theta},\mathbf{\Theta'}\right)  \diff\Delta\tau \diff\Delta f_d}.
\end{equation}
In the following derivation, we assume that all the $K$ waveforms are sharing the same bandwidth and time duration, \colorR{meaning that the integration of the auto-AF for each waveform over region $A$ has a fixed volume $V_0$, i.e., $V_j=V_0, \, \forall j \in \{1,\ldots,K\}$. By substituting \eqref{eq:ambfFuncSeprWood} into \eqref{eq:volumnDef}, the volume of the TB-based MIMO radar AF for the worst-case scenario} can be expressed as
%
%
%
%
%
%
%
%
\begin{alignat}{2}
  V_\mathrm{TB}^\mathrm{I}\left(A\right)  &	\simeq   \frac{E}{K}\left\vert\mathbf{a}_{\mathrm{R}}^H\left(\mathbf{\Theta}\right)\mathbf{a}_{\mathrm{R}} \left(\mathbf{\Theta}'\right)\right\vert^2  \iint_A{\left(\sum^K_{k=1}\left|\Upsilon_k\right|^2\right)}
 {\left|\left[\boldsymbol{\overline{\chi}}\right]_{kk}\left(\Delta\tau,\Delta f_d\right)\right|^2  \diff \Delta\tau \diff \Delta f_d} \nonumber \\
& =  \frac{E}{K}\left\vert\mathbf{a}_{\mathrm{R}}^H\left(\mathbf{\Theta}\right)\mathbf{a}_{\mathrm{R}} \left(\mathbf{\Theta}'\right)\right\vert^2 \left(\sum^K_{k=1}\left|\Upsilon_k\right|^2\right)  V_0  \label{eq:volumeTBI} \\
& \triangleq  V_{K} \label{eq:VKDef}
\end{alignat}
where $\Upsilon_k \triangleq \mathbf{a}_{\mathrm{T}}^H(\boldsymbol\Theta)\mathbf{c}_k, \, k \in \{1,\ldots,K\}$ is the $k$th \colorT{transmit} coherent processing gain that has been defined \colorR{before.}

Employing the Siebert's self-transform property \eqref{eq:SiebertSelfTr} and Parseval's theorem, under the condition that $\psi\left(\tau,f_d\right)$ is any quadratically integrable function whose Fourier transform is
\setcounter{mytempeqncnt2}{33}
\setcounter{equation}{\value{mytempeqncnt2}}
\begin{equation}
\Psi    \left(\tau,f_d\right)   =   \iint_{-\infty}^{\infty}{\psi   \left(\sigma,\nu\right)     \mathrm{exp}\left\{{-j2\pi\nu\tau+j2\pi f_d\sigma}\right\}  \diff \sigma \diff\nu}
\end{equation}
the following transform
%
%
%
%
%
%
%
\newlength{\newIII}
\settowidth{\newIII}{$=   \frac {E}{K}    \iint_{A'} \;\;\;\,$}
\begin{alignat}{3}
  &\iint_A  {\chi\left(\mathbf{\Theta},\mathbf{\Theta'}\right)  \phi\left(\Delta\tau,\Delta f_d\right)  \diff \Delta\tau \diff \Delta f_d}
     \nonumber\\
&
=   \frac {E}{K}
    \iint_{A'}
        {
        \left\vert
                \mathbf{a}_{\mathrm{R}}^H\left(\mathbf{\Theta}\right)
                \mathbf{a}_{\mathrm{R}}\left(\mathbf{\Theta}'\right)
        \right\vert^2}
 \sum_{k=1}^{K}\sum_{j=1}^{K}\left\vert \Upsilon_k\mathbf{a}_{\mathrm{TE}}(j)\right\vert^2 \nonumber \\
  & \hspace*{\newIII} \times \left[\boldsymbol{\xoverline{\chi}}\right]_{kk}^\ast\left(\Delta\tau,\Delta f_d\right) \left[\boldsymbol{\xoverline{\chi}}\right]_{jj}\left(\Delta\tau,\Delta f_d\right)
   \Psi \left(\Delta\tau,\Delta f_d\right)
        d\Delta\tau d\Delta f_d  \qquad \label{eq:TransformHoldI}  \\
        &  \triangleq  V_\mathrm{TB}^\mathrm{I'}\left(A'\right)  \nonumber
\end{alignat}
holds.

Under the assumption that $A$ is convex, symmetric around the origin, and furthermore contains a delta function at the origin, it can be shown using the approach in \cite{PriceBounds} that \colorR{the following inequality}
\begin{alignat}{3}
V_\mathrm{TB}^\mathrm{I}\left(A\right) & >   \frac{1}{4} C\left(A\right)     \underset{A'\to
                                                    0}{\lim}V_\mathrm{TB}^\mathrm{I'}\left(A'\right)\nonumber\\
& =  \frac{1}{4}  C\left(A\right)
                \frac {N^2
                       \left(\sum_{k=1}^{K}\sum_{j=1}^{K}\left\vert \Upsilon_k\mathbf{a}_{\mathrm{TE}}(j)\right\vert^2\right)
                       }
                       {\left\vert
                                \mathbf{a}_{\mathrm{R}}^{\colorL{\text H}}\left(\mathbf{\Theta}\right)
                                \mathbf{a}_{\mathrm{R}}\left(\mathbf{\Theta}'\right)
                        \right\vert^2
                        \left(\sum_{k=1}^{K}\left\vert\Upsilon_k\right\vert^2\right)
                       }
    V_{K}\nonumber\\
& =  \frac{1}{4}  C\left(A\right)
                \frac {N^2
                       K
                       }
                       {\left\vert
                                \mathbf{a}_{\mathrm{R}}^{\colorL{\text H}}\left(\mathbf{\Theta}\right)
                                \mathbf{a}_{\mathrm{R}}\left(\mathbf{\Theta}'\right)
                        \right\vert^2
                       }
    V_{K}\label{eq:VolumnRelation}
\end{alignat}
\colorR{holds,} where $C\left(A\right)$ denotes the area of $A$, \colorR{and $V_K$ is defined in \colorT{\eqref{eq:VKDef}}.}

Based on \eqref{eq:VolumnRelation} and considering the "$\eta$-clear" area that is convex and symmetric around the origin \colorR{with} $\chi\left(\mathbf{\Theta},\mathbf{\Theta'}\right)\leq\eta$, we obtain that the following inequality \colorR{for the the worst-case ``clear region'' of the TB-based MIMO radar AF}
\begin{equation}\label{eq:clearRegion}
\begin{split}
C^{\text{I}}_{\colorR{\text{TB}}}\left(A\right)  \leq   \frac{4V_{K}}{\frac {N^2K
                       }
                       {\left\vert
                                \mathbf{a}_{\mathrm{R}}^{\colorL{\text H}}\left(\mathbf{\Theta}\right)
                                \mathbf{a}_{\mathrm{R}}\left(\mathbf{\Theta}'\right)
                        \right\vert^2
                       }    V_{K} -4\eta}
\end{split}
\end{equation}
which holds if and only if
\begin{equation}
\eta <\frac           {N^2
                       K  V_{K}
                       }
                       {4\left\vert
                        \mathbf{a}_{\mathrm{R}}^{\colorL{\text H}}\left(\mathbf{\Theta}\right)
                        \mathbf{a}_{\mathrm{R}}\left(\mathbf{\Theta}'\right)
                        \right\vert^2
                       }.
\end{equation}

\subsection{\colorR{Best-Case Bound}}
\colorR{In the best-case,} based on the same assumptions for the transmitted waveforms as made in the worst-case, and using also \eqref{eq:ambfFuncSeprWood}, \eqref{eq:volumnDef} can be expressed as
\begin{equation}
\begin{split}
V_\mathrm{TB}^\mathrm{II}\left(A\right)=V_\mathrm{TB}^\mathrm{I}\left(A\right)\simeq  V_{K}.
\end{split}
\end{equation}
Similarly, the following transform 
%
%
%
%
%
%
%
%
\begin{alignat}{2}
 & \iint_A {     \chi\left(\mathbf{\Theta},\mathbf{\Theta'}\right)   \psi\left(\Delta\tau,\Delta f_d\right)  \diff \Delta\tau  \diff \Delta f_d} \nonumber \\
 & =  \frac {E}{K} \iint_{A'}
        {
        \left\vert
                \mathbf{a}_{\mathrm{R}}^H\left(\mathbf{\Theta}\right)
                \mathbf{a}_{\mathrm{R}}\left(\mathbf{\Theta}'\right)
        \right\vert^2}
        \sum_{k=1}^{K}\left\vert\Upsilon_k\right\vert^2
   \left\vert\left[\boldsymbol{\xoverline{\chi}}\right]_{kk}\left(\Delta\tau,\Delta f_d\right)\right\vert^2
   \Psi \left(\Delta\tau,\Delta f_d\right)
        \diff \Delta\tau \diff \Delta f_d  \label{eq:TransformHoldII}  \\
  &   \triangleq  V_\mathrm{TB}^\mathrm{II'}\left(A'\right) \nonumber
\end{alignat}
holds.
\colorR{Under} the same condition as \colorR{applied} in the worst-case, it can be shown that \colorR{the following inequality}
\begin{alignat}{3}
  V_\mathrm{TB}^\mathrm{II}\left(A\right) & >   \frac{1}{4} C\left(A\right)     \underset{A'\to
                                                    0}{\lim}V_\mathrm{TB}^\mathrm{II'}\left(A'\right) \nonumber\\
& =  \frac{1}{4}  C\left(A\right)
                \frac {N^2
                       \left(\sum_{k=1}^{K}\left\vert\Upsilon_k\right\vert^2\right)
                       }
                       {\left\vert
                                \mathbf{a}_{\mathrm{R}}^{\colorL{\text H}}\left(\mathbf{\Theta}\right)
                                \mathbf{a}_{\mathrm{R}}\left(\mathbf{\Theta}'\right)
                        \right\vert^2
                        \left(\sum_{k=1}^{K}\left\vert\Upsilon_k\right\vert^2\right)
                       }
    V_{K} \nonumber\\
& =  \frac{1}{4}  C\left(A\right)
                \frac {N^2
                       }
                       {\left\vert
                                \mathbf{a}_{\mathrm{R}}^{\colorL{\text H}}\left(\mathbf{\Theta}\right)
                                \mathbf{a}_{\mathrm{R}}\left(\mathbf{\Theta}'\right)
                        \right\vert^2
                       }
    V_{K} \label{eq:VolumnRelation2}
\end{alignat}
\colorR{holds.}

Based on \eqref{eq:VolumnRelation2} and considering the "$\eta$-clear" area that is convex and symmetric around the origin for $\chi\left(\mathbf{\Theta},\mathbf{\Theta'}\right)\leq\eta$, we obtain the following inequality \colorR{for the best-case ``clear region'' of the TB-based MIMO radar AF}
\begin{equation}\label{eq:clearRegion2}
\begin{split}
C^{\text{II}}_{\colorR{\text{TB}}}\left(A\right)  \leq   \frac{4V_{K}}{\frac {N^2
                       }
                       {\left\vert
                                \mathbf{a}_{\mathrm{R}}^{\colorL{\text H}}\left(\mathbf{\Theta}\right)
                                \mathbf{a}_{\mathrm{R}}\left(\mathbf{\Theta}'\right)
                        \right\vert^2
                       }    V_{K} -4\eta}
\end{split}
\end{equation}
which holds if and only if
\begin{equation}
\eta <\frac           {N^2
                         V_{K}
                       }
                       {4\left\vert
                        \mathbf{a}_{\mathrm{R}}^{\colorL{\text H}}\left(\mathbf{\Theta}\right)
                        \mathbf{a}_{\mathrm{R}}\left(\mathbf{\Theta}'\right)
                        \right\vert^2
                       }.
\end{equation}

\subsection{\colorR{Discussion}}

\colorR{The worst- and best-case ``clear region'' bounds in \eqref{eq:clearRegion} and \eqref{eq:clearRegion2} which correspond to the two identified limiting cases indicate that they} depend on the array configuration, \colorR{and the quantity $N^2/ |
                            \mathbf{a}_R^{\colorL{\text H}} (\mathbf{\Theta} )
                            \mathbf{a}_R (\mathbf{\Theta}' )
                  |^2$
makes these two bounds} variable. The smaller the quantity is, the larger the maximum possible ``clear region'' bound can be obtained. The largest bound is achieved when this quantity is decreased to 1 \colorR{as long as the $\eta$-level condition is guaranteed}.

The  ``clear region'' bound for the \colorT{worst case} indicates that the worst \colorR{achievable} ``clear region'' \colorR{of} the TB-based MIMO radar AF is independent of the \colorT{transmit} coherent gains, \colorR{however,} it depends on the number of transmitted waveforms $K$ \colorR{under the} condition that the \colorR{emitted waveforms} share the same \colorR{characteristic} parameters and have the same properties. In this sense, it is similar to the case of the traditional MIMO radar AF \colorR{with $K$ mutually orthogonal waveforms that has been given} in \cite{AbramovichBounds}. However, the \colorR{worst-case bound derived here} clarifies that the worst-case \colorR{``clear region'' of the TB-based MIMO radar AF} is inversely proportional to the number of orthogonal waveforms \colorR{$K$ (or the number of beams),} but not the number of transmit antenna elements \colorR{$M$.} Contrarily, the best-case ``clear region'' bound indicates that the ideal ``clear region'' for the TB-based MIMO radar AF is independent of the waveform number \colorR{$K$,} and it is equivalent to the case of the PA radar AF with \colorR{a} single waveform \colorR{that has been} shown in \cite{PriceBounds}.

\colorR{
It is worth noting from analyzing \eqref{eq:volumeTBI} that $V_K$ defined in \eqref{eq:VKDef} is partially determined by the sum of squared magnitudes of the \colorT{transmit} coherent processing gains $\Upsilon_k, \, k=1,\ldots,K$, which means that it is subjected to the TB matrix $\mathbf{C}$ employed by the TB-based MIMO radar system. This quantity, together with the one resulted from the receive array geometry, determines how small the $\eta$-level can be for the TB-based MIMO radar AF. The PA radar and the traditional MIMO radar have their own fixed forms of the TB matrices, \colorT{therefore,} their AFs achieve fixed values of volume $V_K$ under the conditions \eqref{eq:AFCondtions}. Different from the former two, the TB-based MIMO radar uses its own TB matrix $\mathbf{C}$, which makes its maximum ``clear region'' varying in the range bounded by the worst- and best-case bounds. This leads to significant differences between the results achieved for the traditional MIMO radar AF in \cite{AbramovichBounds} and that achieved for the TB-based MIMO radar AF.}

\colorR{
The actual maximum achievable ``clear region'' of the TB-based MIMO radar AF is bounded on both sides by the two identified limiting cases. The worst-case bound becomes larger as $K$ decreases.} Consequently, \colorR{there exists} a tradeoff between the maximum achievable ``clear region'' and the waveform diversity for the TB-based MIMO radar AF. \colorR{Once the desired radar system and target parameters are selected, the TB-based MIMO radar AF can be evaluated directly using its definition \eqref{eq:ambFuncFnl} or simplification \eqref{eq:ambfFuncSeprWood}. This facilitates the radar designer to find the best tradeoff. The worst- and best-case bounds derived in \eqref{eq:clearRegion} and \eqref{eq:clearRegion2} also implicate that the traditional MIMO radar AF achieves the worst maximum achievable ``clear region'', and it is approximately $1/M$  that of the PA radar, which agrees with the result of \cite{AbramovichBounds}. It is clear that the maximum achievable ``clear region'' of the TB-based MIMO radar AF is in between that of the PA and traditional MIMO radar cases.}

\colorH{In reality, orthogonal waveforms for all time delays and Doppler shifts do not exist \colorT{\cite{ForsytheMIMOWavCon10}.} Indeed, the TB-based MIMO radar makes it relatively easier to achieve waveforms with better orthogonality because fewer waveforms \colorH{are} needed, provided that the degrees of freedom are enough. The above ``clear region'' analysis is essentially the way to investigate the non-orthogonal effects of the $K$ original waveforms employed in the TB-based MIMO radar, which is highly related to sidelobe analysis. There exist waveform design methods based on minimizing or explicitly constraining the sidelobe levels of the transmitted waveforms \cite{ChenAFWaveform08}, \cite{BlumWaveform07}, \cite{LiSignalSyn} and other design methods such as the ones based on time/Doppler division \cite{ForsytheMIMOWavCon10} or space-time coding \cite{DengPolyphaseWaveform, RabideauWaveform08, SongWaveform10}.} Hence, large ``clear region'' under the ``$\eta$-clear'' condition can be achieved. To further obtain a larger ``clear region'' for the TB-based MIMO radar AF, one can resort to the range-Doppler sidelobes mitigation techniques. For example, receiver instrumental variable filter \cite{LiSignalSyn, MaSidelobeSuppress, HuaRangeDopplerSupp} can be employed at the receiving end to suppress the sidelobes. However, the attainable ``clear region'' depends on the exact sidelobe mitigation level.

\section{Simulation Results}\label{Sec:Simulation}

In this section, we provide numerical examples to demonstrate the AFs for different radar configurations using the generalized TB-based MIMO radar AF definition given in this paper. We also present comparisons to the AF metric of \cite{AbramovichBounds}.

Throughout \colorR{the} simulations, we assume \colorR{that uniform linear arrays of $M=8$ omni-directional transmit antenna elements and $N=8$ omni-directional receive antenna elements spaced half a wavelength apart from each other are used.} Both the transmit and receive arrays are located on the $x$-axis \colorR{with their centers being located at the origin.} The total transmit energy \colorR{$E$ is fixed to be equal to the number of transmit antenna elements $M$.} Optimized polyphase-coded waveforms \cite{LevanonRadarSignal} of code length 512 are used. We employ a single pulse for simulating the AF, and the pulse width \colorL{$T_p$} is selected to be \colorH{$60 \, \mu \text{s}$.} The time-bandwidth product \colorR{$B\colorL{T_p}$} is set to be \colorR{equal to 256}, and the sampling rate \colorR{$f_s$} is \colorR{selected to be two times of the bandwidth, i.e., $f_s=2B$.} 
We fix both target parameters $\boldsymbol{\Theta}$ (with zero Doppler) and $\boldsymbol{\Theta}'$ in the $x$-$y$ plane, and the latter is varying. We show all together four examples, and in the first two examples, both parameters are set to share the same spatial angle $\theta=0^\circ$. While in the last two examples, both parameters are set to share the same delay $\tau=0$, but $\boldsymbol{\Theta}'$ is allowed to have different spatial angles. The \colorR{maximum} magnitudes of all the simulated AFs are normalized to \colorR{1}, thus, the mainlobes of all simulated AFs are 0 dB. 

For the TB-based MIMO radar configuration, we select the spatial sector-of-interest as $\boldsymbol{\Omega}=[-15^\circ,15^\circ]$ with a $10^\circ$ wide transition band on each side. The transmit energy is focused  within $\boldsymbol{\Omega}$ and it is radiated via $K=4$ transmit beams. Two TB design strategies are used in the simulations. One is the TB design \eqref{eq:TBAFConvex} which is used in the second and third examples, and the other is the TB design \eqref{eq:TBAFConvexNew} which is used in the first and last examples. For the TB design \eqref{eq:TBAFConvex}, the desired vector $\mathbf{d}(\theta)$ is selected as\footnote{For such $\mathbf{d}(\theta)$, the rotational invariance property (RIP) \cite{OptimumArrayPro, XuBeamspaceESPRIT} is guaranteed that may be desirable in DOA estimation applications \cite{VorobyovTBDesign, VorobyovTRBeamDesign}.} $\mathbf{d}(\theta) \triangleq [\mathrm{exp}\{\mu_1(\theta)\}, \ldots, $ $\mathrm{exp}\{\mu_4(\theta)\}]^{\colorL{\text T}}$ where $\mu_k(\theta), \, k \in\{1,\ldots,4\}$ is the $k$th linear function of the spatial angle $\theta$. The parameter $\gamma$ of this design that controls the level of radiated power outside $\boldsymbol{\Omega}$ is selected differently in the second and third examples. For the TB design \eqref{eq:TBAFConvexNew}, the desired vector $\mathbf{d} \left( \theta \right)$ is selected as the same in \eqref{eq:TBAFConvex}, and the Doppler domain $\Delta f_d = [-30 \, \mathrm{kHz}, -18 \, \mathrm{kHz}] \cup [18 \, \mathrm{kHz}, 30 \, \mathrm{kHz}]$ at the spatial direction of $\theta=0^\circ$ is selected to be controlled. Other parameters of this design are respectively selected as $\gamma=0.1$, $\delta=0.3$, $\boldsymbol\vartheta=\mathbf{0}^\circ$, and $f_d^0=0 \, \mathrm{kHz}$. We use the CVX MATLAB toolbox \cite{CVX} to solve both TB optimization design problems in the simulations.

\begin{figure}[H]
	\centering
    \subfloat[\label{subfig3:Fig-a}] { \includegraphics[width=0.6\textwidth]{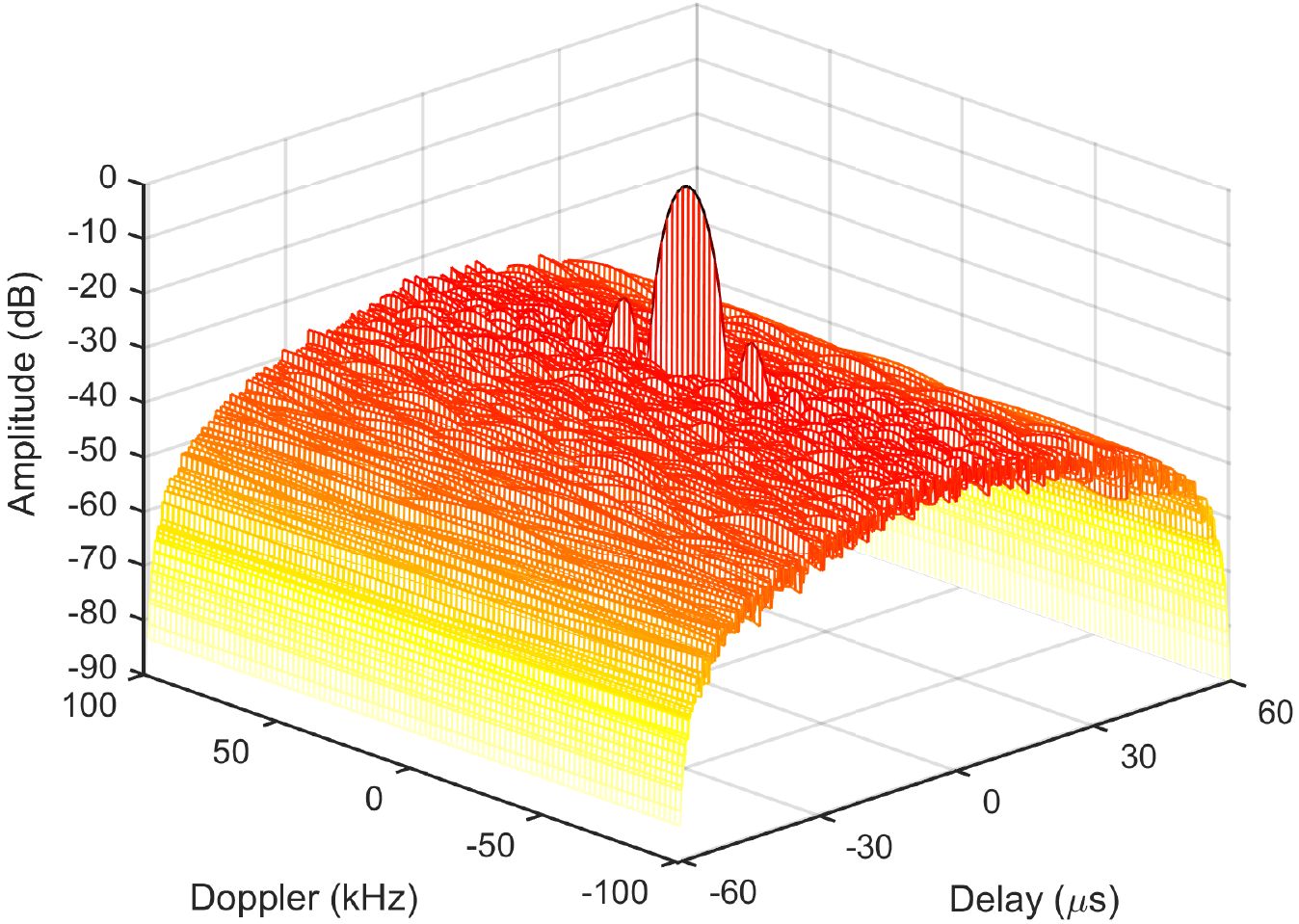} }
    \vfill
    \subfloat[\label{subfig3:Fig-b}] {\includegraphics[width=0.6\textwidth]{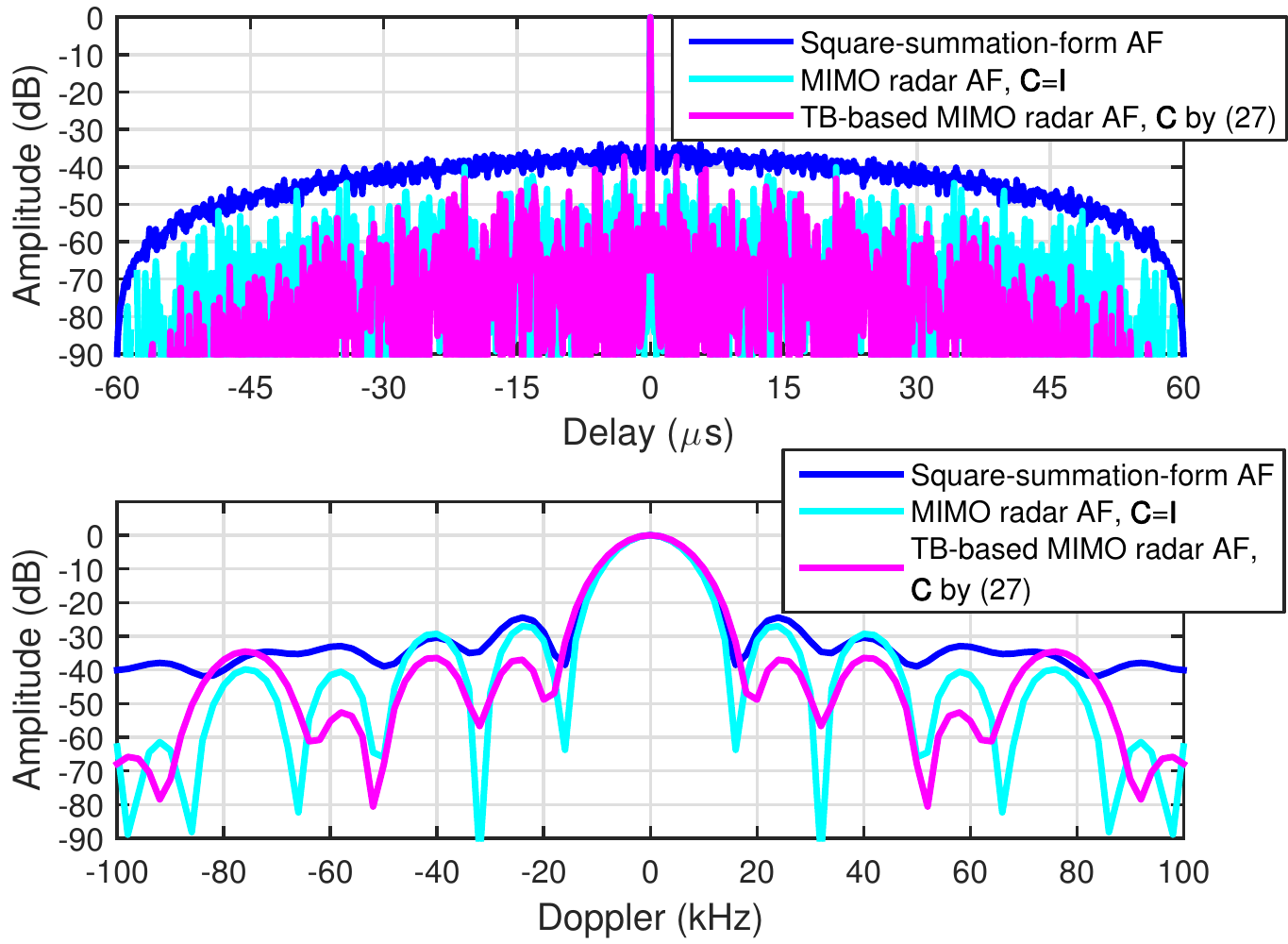} }
    \caption{Comparison to the square-summation-form MIMO radar AF of \cite{AbramovichBounds}. Here $M=8$, $N=8$, and $E=M$. (a)~3D view of the square-summation-form MIMO radar AF and (b)~Zero-delay and zero-Doppler cuts of the square-summation-form MIMO radar AF, the MIMO radar AF for $\mathbf{C} = \mathbf{I}$, and the TB-based MIMO radar AF for the proposed design of $\mathbf{C}$ by \eqref{eq:TBAFConvexNew}. The total $K=8$ single-pulse polyphase-coded waveforms of code length 512 are used for the square-summation-form MIMO radar AF and the MIMO radar AF for $\mathbf{C}=\mathbf{I}$, and the first $K=4$ waveforms are used for the TB-based MIMO radar AF for design of $\mathbf{C}$ by \eqref{eq:TBAFConvexNew}: $\colorL{T_p}=60 \, \mu\text{s}$, $BT_p=256$, and $f_s=2B$. High relative sidelobe levels are obtained in Doppler-delay domain using the square-summation-form AF, and they are highly concentrated and of arched shape over the delay domain (e.g., the range $[-42 \, \text{dB}, -32 \, \text{dB}]$ of AF amplitudes versus the delay interval $[-15 \, \mu \text{s}, 15 \, \mu \text{s}]$). Dispersive relative sidelobes (with nulls) are obtained using the latter two AFs, and their levels are lower than that of the square-summation-form AF. The MIMO radar AF for $\mathbf{C} = \mathbf{I}$ boils down to the existing MIMO radar AF defined in \cite{AntonioMIMOAbmF} because no TB processing is applied. The relative sidelobe levels of the sub Doppler domain $[-30 \, \mathrm{kHz}, -18 \, \mathrm{kHz}] \cup[18 \, \mathrm{kHz}, 30 \, \mathrm{kHz}]$ is controlled to approach $-40$~dB for the TB-based MIMO radar AF for design of $\mathbf{C}$ by \eqref{eq:TBAFConvexNew}. The square-summation-form AF metric of \cite{AbramovichBounds} shows the worst ``clear region''.} 
    \label{Fig:FigMIMOAFCaseI}
\end{figure}

In the first example, we show the square-summation-form MIMO radar AF metric defined in \cite{AbramovichBounds} (see Fig.~\ref{Fig:FigMIMOAFCaseI}\subref{subfig3:Fig-a}), and compare its zero-delay and zero-Doppler cuts to that of the MIMO radar AF for $\mathbf{C} = \mathbf{I}$ and the TB-based MIMO radar AF for the proposed design of $\mathbf{C}$ by \eqref{eq:TBAFConvexNew} (see Fig.~\ref{Fig:FigMIMOAFCaseI}\subref{subfig3:Fig-b}). It can be seen from Fig.~\ref{Fig:FigMIMOAFCaseI} that the relative sidelobe levels of the square-summation-form AF are highly concentrated and of arched shape over the delay domain. For example, it can be viewed from delay domain that all AF sidelobes in the range of delays within $[-15 \, \mu \text{s}, 15 \, \mu \text{s}]$ are above $-42$~dB, and the highest level of sidelobes around the AF mainlobe reaches approximately $-32$~dB. From the comparison result of zero-delay and zero-Doppler cuts of AF, it can be seen from Fig.~\ref{Fig:FigMIMOAFCaseI}\subref{subfig3:Fig-b} that dispersive relative sidelobes (with nulls) are obtained using the latter two AFs, and their levels are lower than that of the square-summation-form AF. The MIMO radar AF for $\mathbf{C}=\mathbf{I}$ boils down to the MIMO radar AF defined in \cite{AntonioMIMOAbmF} because no TB processing is applied. To maintain good Doppler tolerance, we can enforce spectral constraints \cite{RoweSpecConstraint14} in addition to ensuring good waveform correlation (i.e., zero-Doppler cut of AF) property when designing the waveforms. For the TB-based MIMO radar, we also have the choice to control Doppler sidelobe levels by designing the TB matrix. 
It can also be seen from Fig.~\ref{Fig:FigMIMOAFCaseI}\subref{subfig3:Fig-b} that the Doppler relative sidelobe levels are suppressed to about $-40$~dB. Indeed, this example implicates that the square-summation-form AF metric obtains worse ``clear region'' than that of the AF defined in this paper for a given allowable sidelobe level limit $\eta$. In other words, the sidelobe level limit for the AF in \cite{AbramovichBounds} can only be set to a relatively high value.

\begin{figure}[H]
	\centering
    \subfloat[\label{subfig4:Fig-b}] { \includegraphics[width=0.7\textwidth]{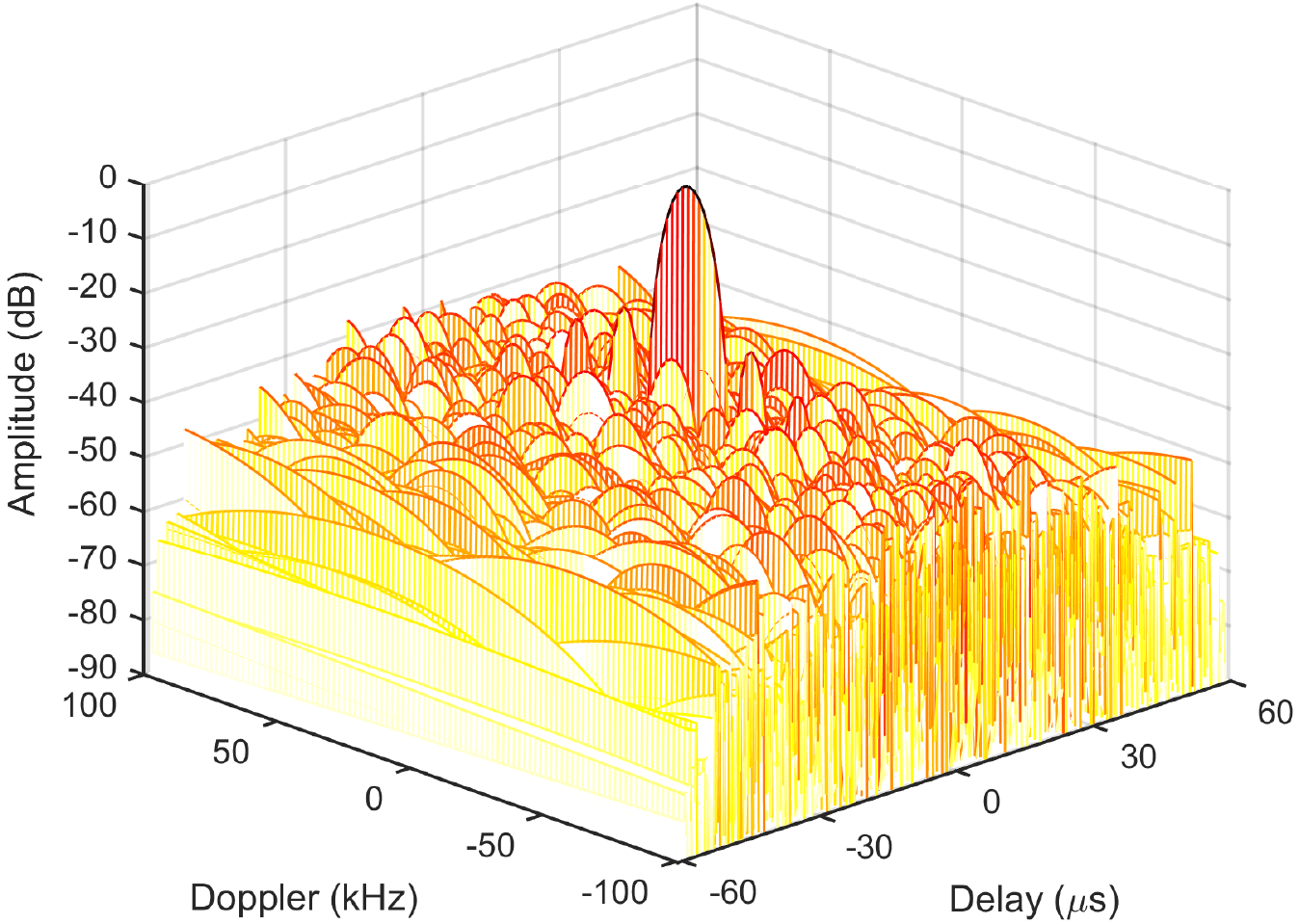} }
    \vfill
    \subfloat[\label{subfig4:Fig-a}] { \includegraphics[width=0.7\textwidth]{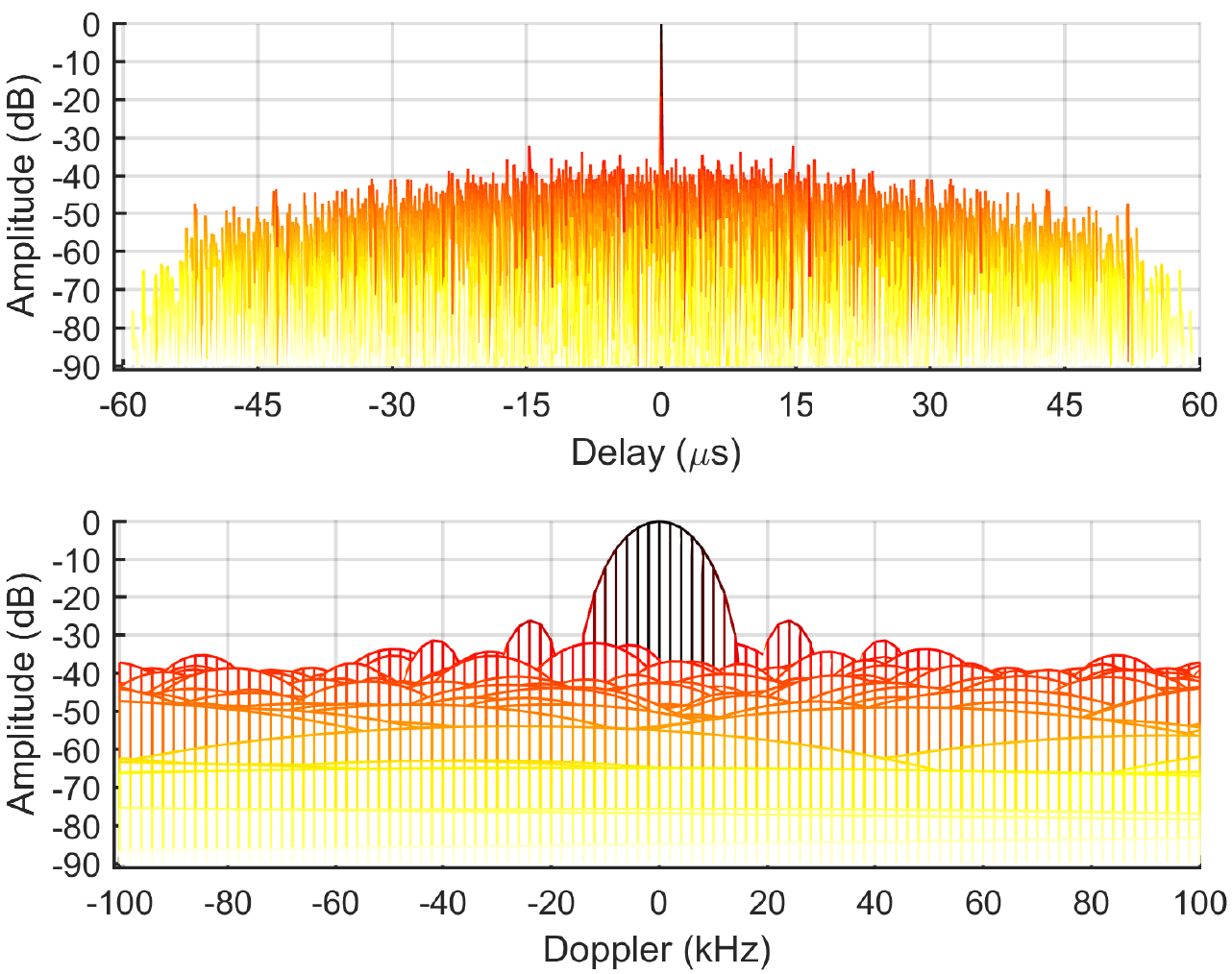} }
    \caption{The TB-based MIMO radar AF (versus delays and Doppler) associated with the TB design \eqref{eq:TBAFConvex}. Here $M=8$, $N=8$, \colorR{$K=4$,} and $E=M$. The first $4$ out of $8$ single-pulse polyphase-coded waveforms of code length 512 are used: $T_p=60 \, \mu\text{s}$, $BT_p=256$, and $f_s=2B$. (a)~3D view of AF and (b)~2D views of the AF.  Low relative sidelobe levels are achieved in Doppler-delay domain using this AF, and they are dispersive. Dispersive relative sidelobes are obtained and the average relative sidelobe levels can be seen lower than that present in the previous example. Doppler relative sidelobe levels are high because no control is implemented.} 
    \label{Fig:FigTBMIMOAFConvexI}
    \vspace*{-4pt}
\end{figure}

In the second example, we show the TB-based MIMO radar AF versus delays and Doppler (see Fig.~\ref{Fig:FigTBMIMOAFConvexI}) where the TB deisgn strategy \eqref{eq:TBAFConvex} is employed. The target velocity is not needed when carrying out the optimization process, thus we employ the spatial angle $\theta$ to replace $\boldsymbol{\Theta}$ in all the steering vectors. The parameter $\gamma$ is selected to be $0.38$. The figure shows the 3D (full view) result in the first subfigure, and the 2D (side view) result in the second subfigure. It can be seen \colorR{from both subfigures} that the relative sidelobes of the TB-based MIMO radar AF are dispersive. From the view of delay domain, it can be seen that the major sidelobes around the AF mainlobe concentrate on the level of $-50$~dB. While from the view of Doppler domain, it can be seen that the average level of major sidelobes is about $-40$~dB. The worst sidelobe level from this view is about $-27$~dB, which is because the convex optimization based TB design \eqref{eq:TBAFConvex} does not consider the factor of Doppler processing.

\begin{figure}[!t]
	\centering
    \subfloat[\label{subfig5:Fig-b}] { \includegraphics[width=0.7\textwidth]{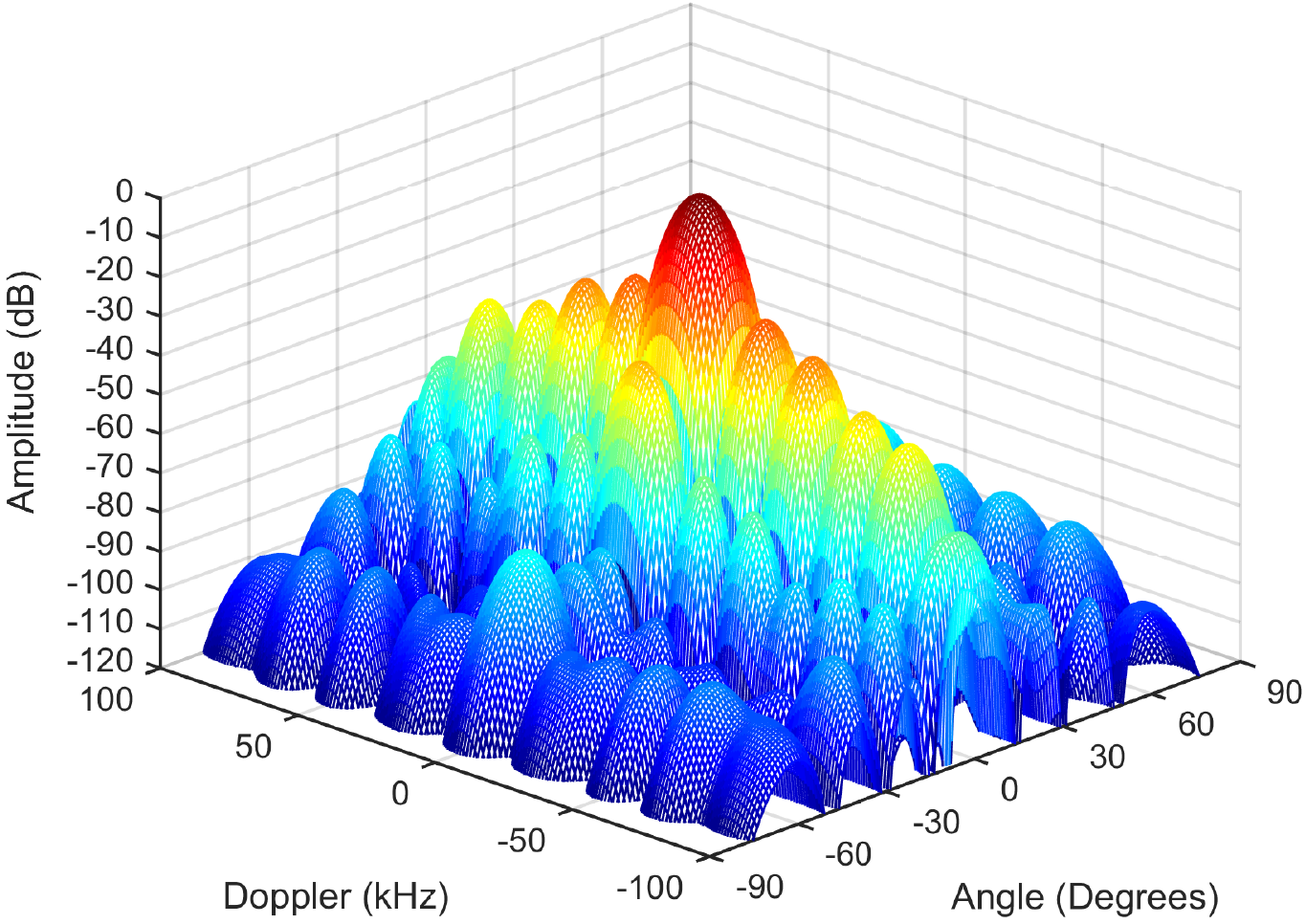} }
    \vfill
    \subfloat[\label{subfig5:Fig-a}] { \includegraphics[width=0.7\textwidth]{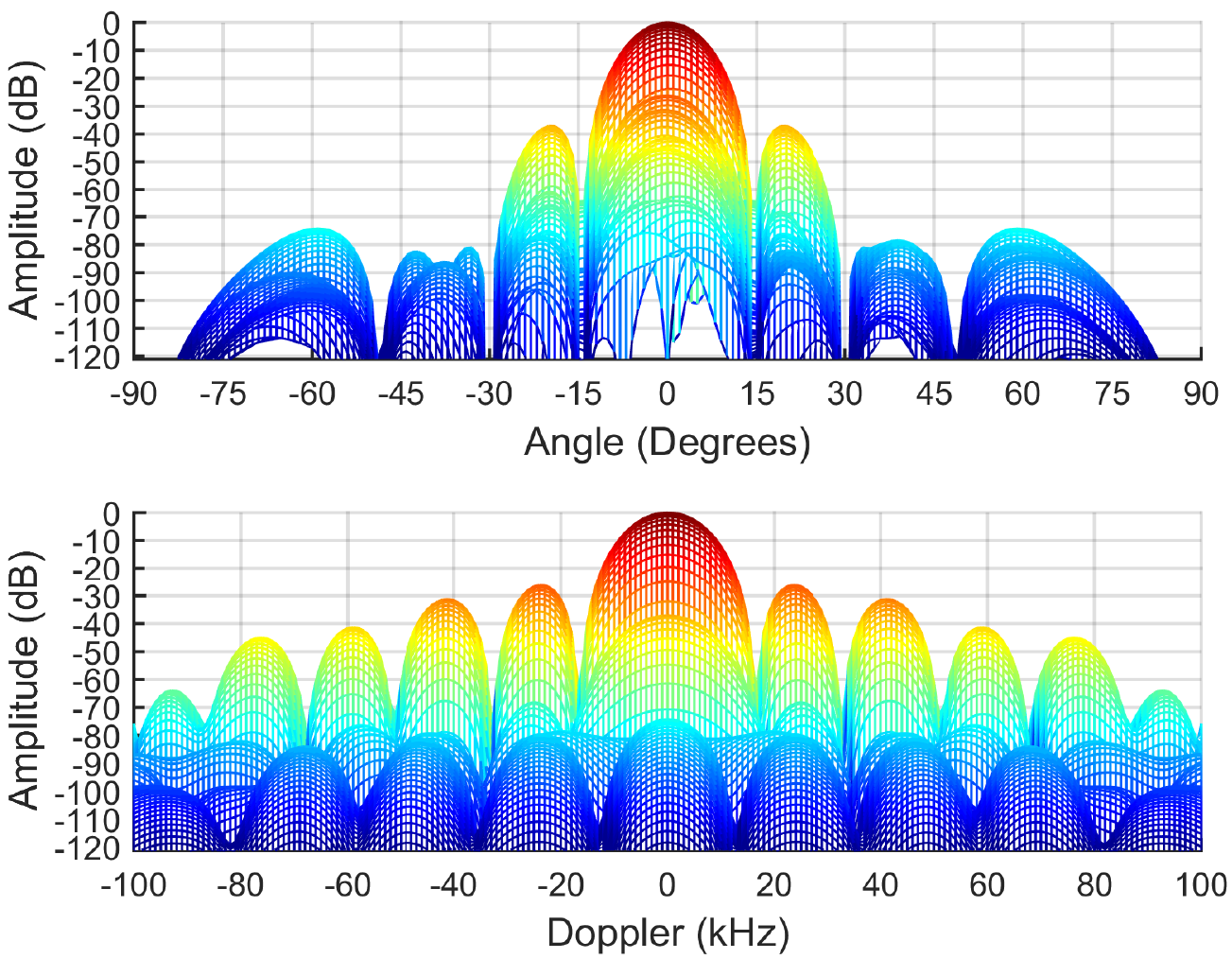} }
    \caption{The TB-based MIMO radar AF (versus angles and Doppler) associated with the TB design \eqref{eq:TBAFConvex}. Here $M=8$, $N=8$, \colorR{$K=4$,} and $E=M$. The first 4 single-pulse polyphase-coded waveforms of code length 512 are used: $T_p=60 \, \mu\text{s}$, $B\colorL{T_p}=256$, and $f_s=2B$. (a)~3D view of AF and (b)~2D views of the AF. The information of transmit energy focusing within the spatial sector $[-15^\circ, 15^\circ]$ is conveyed by the result. Low relative sidelobe levels of the AF are achieved in Doppler-angle domain. Due to the reason that the used TB design does not consider the effect of Doppler processing, some relative sidelobe levels along the spatial direction $\theta=0^\circ$ are still high.}
    \label{Fig:FigTBMIMOAFConvexII}
\end{figure}

\colorR{In the third example, we show the TB-based MIMO radar AF versus Doppler and spatial angles (see Fig.~\ref{Fig:FigTBMIMOAFConvexII}). The convex optimization based strategy \eqref{eq:TBAFConvex} is used to design $\mathbf{C}$. All other simulation parameters are the same as in the previous example except the parameter $\gamma$ which is selected as $0.2$. To better display the results, we remove all the sidelobes that are below $-120$~dB. The 3D and 2D results are shown in the first and second subfigures, respectively. It can be seen from the 3D subfigure that the TB-based MIMO radar AF has lower sidelobe levels versus angles than that versus Doppler. From the view of angle, the AF in fact shows the beampattern of the TB-based MIMO radar system, and the highest relative sidelobe level in this view is about $-28$~dB. From the view of Doppler, the worst relative sidelobe level reaches about $-27$~dB. As the same in the previous example, this level is relatively high due to the reason that the design of the TB matrix does not consider the factor of Doppler processing.
}

\begin{figure}[!t]
	\centering
    \subfloat[\label{subfig6:Fig-b}] { \includegraphics[width=0.7\textwidth]{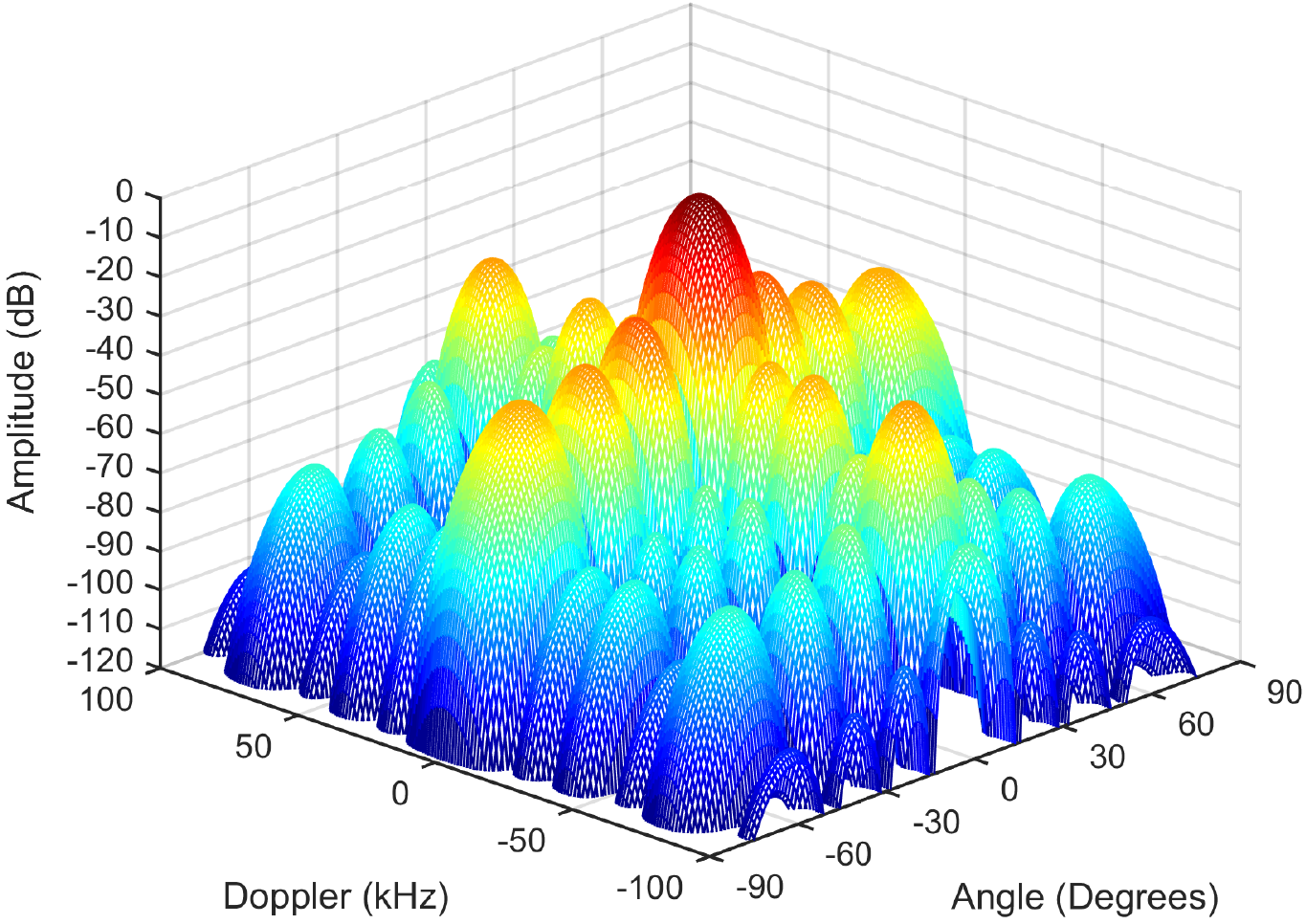} }
    \vfill
    \subfloat[\label{subfig6:Fig-a}] { \includegraphics[width=0.7\textwidth]{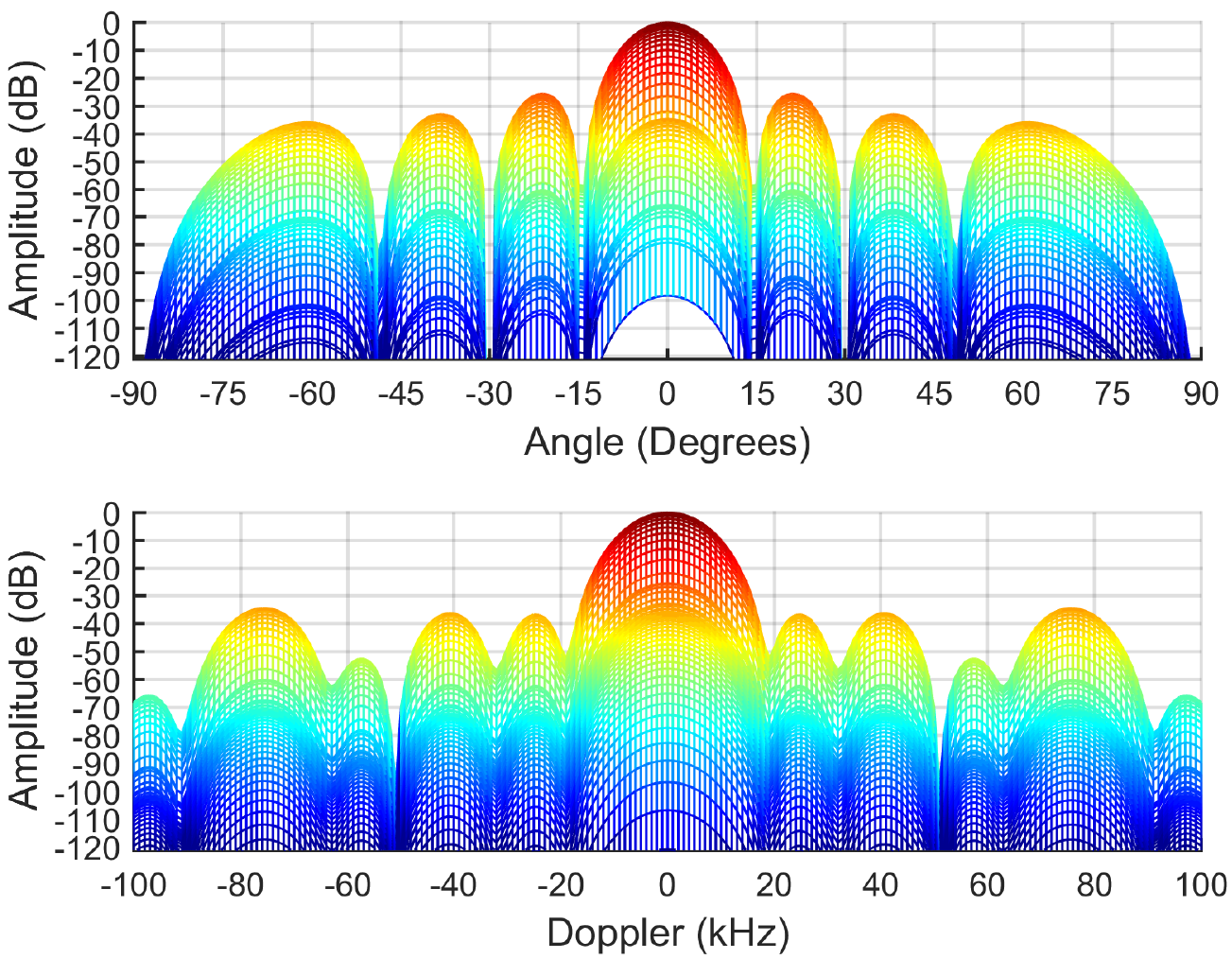} }
    \caption{The TB-based MIMO radar AF (versus angles and Doppler) associated with the proposed TB design \eqref{eq:TBAFConvexNew}. Here $M=8$, $N=8$, \colorR{$K=4$,} and $E=M$. The first 4 single-pulse polyphase-coded waveforms are used: $T_p=60 \, \mu\text{s}$, $B\colorL{T_p}=128$, and $f_s=2B$. (a)~3D view of AF and (b)~2D views of the AF. The relative sidelobe levels of AF in sub Doppler domain $[-30 \, \mathrm{kHz}, -18 \, \mathrm{kHz}] \cup$ $[18 \, \mathrm{kHz}, 30 \, \mathrm{kHz}]$ along the spatial direction $\theta=0^\circ$ are artificially controlled by the proposed TB design, and they are well suppressed to below $-38 \, \text{dB}$ which in fact demonstrates the tradeoff between Doppler and spatial processing in the TB-based MIMO radar.}
    \label{Fig:FigTBMIMOAFConvexIII}
\end{figure}

\colorR{In the last example, we show the TB-based MIMO radar AF versus Doppler and spatial angles using the proposed TB design strategy \eqref{eq:TBAFConvexNew} (see Fig.~\ref{Fig:FigTBMIMOAFConvexIII}). We aim at suppressing the relative Doppler sidelobe levels of AF in the range $[-30 \, \mathrm{kHz}, -18 \, \mathrm{kHz}] \cup$ $[18 \, \mathrm{kHz}, 30 \, \mathrm{kHz}]$ at the spatial direction of $\theta=0^\circ$. To better display the results, we remove the sidelobes that are below $-120$~dB. It can be seen from the 2D subfigure that the worst sidelobe level in the desired Doppler range is well suppressed to below $-38$~dB, and the worst sidelobe level in the whole Doppler domain which is far away from the AF mainlobe is about $-35$~dB. Because there is no constraint on the sidelobe levels versus other angles except $\theta=0^\circ$ in \eqref{eq:TBAFConvexNew}, the worst sidelobe level in the whole spatial domain increases to about $-26$~dB. The result shown in this example, indeed, is the tradeoff between Doppler and spatial processing for the TB-based MIMO radar AF, and it newly verifies the validity of tradeoffs in the TB-based MIMO radar \cite{VorobyovSAMTutorial}.}

\section{Conclusion}\label{Sec:Conclu}

\colorR{In this paper, we have derived the AF for the recently proposed TB-based MIMO radar which \colorT{allows obtaining} waveform diversity and \colorT{transmit} coherent processing gains over a pre-defined angular sector simultaneously. Our definition is very general and contains the AFs for the PA, traditional MIMO, and TB-based MIMO radars as important special cases under the standard assumption of far-field targets \colorT{and} narrow-band waveforms.} Relationships among the TB-based MIMO radar AF defined in this paper and the previous \colorR{AF works} including the Woodward's AF, the AF defined for the traditional \colorR{colocated} MIMO radar, \colorR{and the PA radar AF, have been} established, respectively. \colorR{We have compared our newly defined TB-based MIMO radar AF with the existing traditional MIMO radar AFs, and have proposed a new TB design in order to give better relative AF sidelobe levels.} Two \colorR{limiting} cases are identified to bound the ``clear region'' of the TB-based MIMO radar AF, \colorR{and corresponding bounds for these two cases have been derived, respectively. We have shown that the ``clear region'' for the the worst bounding case is inversely proportional to the number of transmitted waveforms $K$, while the one for the best bounding case is independent of $K$. The ``clear region'' of the TB-based MIMO radar AF, which depends on the array configuration, is in between that of the \colorT{worst and best cases.} We have shown in the simulation results that the square-summation-form AF leads to higher relative AF sidelobe levels than that of the TB-based MIMO radar AF. Moreover, using the proposed convex optimization TB design, the levels can be further reduced.}


\begin{thebibliography}{10}
\providecommand{\url}[1]{#1}
\csname url@samestyle\endcsname
\providecommand{\newblock}{\relax}
\providecommand{\bibinfo}[2]{#2}
\providecommand{\BIBentrySTDinterwordspacing}{\spaceskip=0pt\relax}
\providecommand{\BIBentryALTinterwordstretchfactor}{4}
\providecommand{\BIBentryALTinterwordspacing}{\spaceskip=\fontdimen2\font plus
\BIBentryALTinterwordstretchfactor\fontdimen3\font minus
  \fontdimen4\font\relax}
\providecommand{\BIBforeignlanguage}[2]{{%
\expandafter\ifx\csname l@#1\endcsname\relax
\typeout{** WARNING: IEEEtran.bst: No hyphenation pattern has been}%
\typeout{** loaded for the language `#1'. Using the pattern for}%
\typeout{** the default language instead.}%
\else
\language=\csname l@#1\endcsname
\fi
#2}}
\providecommand{\BIBdecl}{\relax}
\BIBdecl

\bibitem{LiMIMORadar}
J.~Li and P.~Stoica, \emph{{MIMO} Radar Signal Processing}.\hskip 1em plus
  0.5em minus 0.4em\relax New York: Wiley, 2009.

\bibitem{HaimovichMIMO}
A.~M. Haimovich, R.~S. Blum, and L.~J. Cimini, ``{MIMO} radar with widely
  separated antennas,'' \emph{{IEEE} Signal Process. Mag.}, vol.~24, no.~1, pp.
  116--129, Jan. 2008.

\bibitem{LiMIMOMag.}
J.~Li and P.~Stoica, ``{MIMO} radar with colocated antennas,'' \emph{{IEEE}
  Signal Process. Mag.}, vol.~24, no.~5, pp. 106--114, Sep. 2007.

\bibitem{FishlerSpatial}
E.~Fishler, A.~Haimovich, R.~S. Blum, L.~Cimini, D.~Chizhik, and R.~Valenzuela,
  ``Spatial diversity in radars---{M}odels and detection performance,''
  \emph{{IEEE} Trans. Signal Process.}, vol.~54, no.~3, pp. 823--838, Mar.
  2006.

\bibitem{FishlerMIMOIdea}
E.~Fishler, A.~Haimovich, R.~S. Blum, D.~Chizhik, L.~Cimini, and R.~Valenzuela,
  ``{MIMO} radar: An idea whose time has come,'' in \emph{Proc. {IEEE} Radar
  Conf.}, Honolulu, HI, Apr. 2004, pp. 71--78.

\bibitem{BlissMIMO03}
D.~W. Bliss and K.~W. Forsythe, ``Multiple-input multiple-output {MIMO} radar
  and imaging: Degrees of freedom and resolution,'' in \emph{Proc. Asilomar
  Conf. Signals, Syst., Comput.}, vol.~1, Pacific Grove, CA, Nov. 2003, pp.
  54--59.

\bibitem{BekkermanTargetDetec06}
I.~Bekkerman and J.~Tabrikian, ``Target detection and localization using {MIMO}
  radar and sonars,'' \emph{{IEEE} Trans. Signal Process.}, vol.~54, no.~10,
  pp. 3873--3883, Oct. 2006.

\bibitem{AbramovichMIMOLimit10}
Y.~I. Abramovich, G.~J. Frazer, and B.~A. Johnson, ``Noncausal adaptive spatial
  clutter mitigation in monostatic {MIMO} radar: {F}undamental limitations,''
  \emph{{IEEE} J. Sel. Topics Signal Process.}, vol.~4, no.~1, pp. 40--54, Feb.
  2010.

\bibitem{LiMIMOSARImaging07}
J.~Li, X.~Zheng, and P.~Stoica, ``{MIMO SAR} imaging: {S}ignal synthesis and
  receiver design,'' in \emph{Proc. {IEEE} Int. Workshop. on Computational
  Advances in Multi-Sensor Adaptive Processing}, St. Thomas, VI, Dec. 2007, pp.
  89--92.

\bibitem{KriegerMIMOSAR14}
G.~Krieger, ``{MIMO-SAR}: {O}pportunities and pitfalls,'' \emph{{IEEE} Trans.
  Geosci. Remote Sens.}, vol.~52, no.~5, pp. 2628--2645, May 2014.

\bibitem{YongzheJamICASSP}
Y.~Li, S.~A. Vorobyov, and A.~Hassanien, ``{MIMO} radar capability on powerful
  jammers suppression,'' in \emph{Proc. {IEEE} Int. Conf. Acoust., Speech,,
  Signal Process. (ICASSP)}, Florence, Italy, May 2014, pp. 5277--5281.

\bibitem{YongzheRobustBeam}
Y.~Li, S.~A. Vorobyov, and A.~Hassanien, ``Robust beamforming for jammers
  suppression in {MIMO} radar,'' in \emph{Proc. {IEEE} Radar Conf.},
  Cincinnati, OH, May 2014, pp. 0629--0634.

\bibitem{FuhrmannHybrid}
D.~R. Fuhrmann, J.~P. Browning, and M.~Rangaswamy, ``Signaling strategies for
  the hybrid {MIMO} phased-array radar,'' \emph{{IEEE} J. Sel. Topics Signal
  Process.}, vol.~4, no.~1, pp. 66--78, Sep. 2010.

\bibitem{VorobyovTransmit}
A.~Hassanien and S.~A. Vorobyov, ``Transmit energy focusing for {DOA}
  estimation in {MIMO} radar with colocated antennas,'' \emph{{IEEE} Trans.
  Signal Process.}, vol.~59, no.~6, pp. 2669--2682, Jun. 2011.

\bibitem{FuhrmannTrasnBeam}
D.~R. Fuhrmann and G.~{San Antonio}, ``Transmit bemforming for {MIMO} radar
  systems using signal cross-correlation,'' \emph{{IEEE} Trans. Aerosp.
  Electron. Syst.}, vol.~44, no.~1, pp. 171--186, Jan. 2008.

\bibitem{StoicaProbingDesing}
P.~Stoica, J.~Li, and Y.~Xie, ``On probing signal design for {MIMO} radar,''
  \emph{{IEEE} Trans. Signal Process.}, vol.~55, no.~8, pp. 4151--4161, Aug.
  2007.

\bibitem{VorobyovPMIMO}
A.~Hassanien and S.~A. Vorobyov, ``Phased-{MIMO} radar: A tradeoff between
  phased-array and {MIMO} radars,'' \emph{{IEEE} Trans. Signal Process.},
  vol.~58, no.~6, pp. 3137--3151, Jun. 2010.

\bibitem{VorobyovTBDesign}
A.~Khabbazibasmenj, A.~Hassanien, S.~A. Vorobyov, and M.~W. Morency,
  ``Efficient transmit beamspace design for search-free based {DOA} estimation
  in {MIMO} radar,'' \emph{{IEEE} Trans. Signal Process.}, vol.~62, no.~6, pp.
  1490--1500, Mar. 2014.

\bibitem{VorobyovTRBeamDesign}
A.~Khabbazibasmenj, S.~A. Vorobyov, A.~Hassanien, and M.~W. Morency, ``Transmit
  beamspace design for direction finding in colocated {MIMO} radar with
  arbitrary receive array and even number of waveforms,'' in \emph{Proc.
  Asilomar Conf. Signals, Syst., and Comput.}, Pacific Grove, CA, Nov. 2012,
  pp. 1307--1311.

\bibitem{VbyovBeamDesignICASSP}
A.~Khabbazibasmenj, S.~A. Vorobyov, and A.~Hassanien, ``Transmit beamspace
  design for direction finding in colocated {MIMO} radar with arbitrary receive
  array,'' in \emph{Proc. {IEEE} Int. Conf. Acoust., Speech,, Signal Process.
  (ICASSP)}, Prague, Czech Republic, May 2011, pp. 2784--2787.

\bibitem{AittomakiLowComTB07}
T.~Aittomaki and V.~Koivunen, ``Low-complexity method for transmit beamforming
  in {MIMO} radars,'' in \emph{Proc. {IEEE} Int. Conf. Acoust., Speech,, Signal
  Process. (ICASSP)}, vol.~II, Honolulu, HI, Apr. 2007, pp. 305--308.

\bibitem{AittomakiCovTB07}
T.~Aittomaki and V.~Koivunen, ``Signal covariance matrix optimization for
  transmit beamforming in {MIMO} radars,'' in \emph{Proc. Asilomar Conf.
  Signals, Syst., Comput.}, Pacific Grove, CA, Nov. 2007, pp. 182--186.

\bibitem{SchmidtDOA}
R.~O. Schmidt, ``Multiple emitter location and signal parameter estimation,''
  \emph{{IEEE} Trans. Antennas Propag.}, vol.~34, no.~3, pp. 276--280, Mar.
  1986.

\bibitem{RoyESPRIT}
R.~Roy and T.~Kailath, ``{ESPRIT}---{E}stimation of signal parameters via
  rotational invariance techniques,'' \emph{{IEEE} Trans. Acoust., Speech,
  Signal Process.}, vol.~37, no.~7, pp. 984--995, Jul. 1989.

\bibitem{PriceBounds}
R.~Price and E.~Hofstetter, ``Bounds on the volume and height distributions of
  the ambiguity function,'' \emph{{IEEE} Trans. Inf. Theory}, vol.~11, no.~2,
  pp. 207--214, Apr. 1965.

\bibitem{SielertSelfTr}
W.~M. Siebert, ``Studies of {W}oodward's uncertainty function,'' Electronics
  Research Lab., M.I.T, Cambridge Massachusetts, MA, Quarterly Progress Rep.,
  Apr. 1958.

\bibitem{AFSLooking65}
R.~O. Harger, ``An optimum design of ambiguity function, antenna pattern, and
  signal for side-looking radars,'' \emph{{IEEE} Trans. Mil. Electron.},
  vol.~9, no.~3, pp. 264--278, Jul. 1965.

\bibitem{AntonioMIMOAbmF}
G.~{San Antonio}, D.~R. Fuhrmann, and F.~C. Robey, ``{MIMO} radar ambiguity
  functions,'' \emph{{IEEE} J. Sel. Topics Signal Process.}, vol.~1, no.~1, pp.
  167--177, Jun. 2007.

\bibitem{ChenAFWaveform08}
C.-Y. Chen and P.~P. Vaidyanathan, ``{MIMO} radar ambiguity properties and
  optimization using frequency-hopping waveforms,'' \emph{{IEEE} Trans. Signal
  Process.}, vol.~56, no.~12, pp. 5926--5936, Dec. 2008.

\bibitem{AbramovichBounds}
Y.~I. Abramovich and G.~J. Frazer, ``Bounds on the volume and height
  distributions for the {MIMO} radar ambiguity function,'' \emph{{IEEE} Signal
  Process. Lett.}, vol.~15, pp. 505--508, 2008.

\bibitem{YongzheAFICASSP}
Y.~Li, S.~A. Vorobyov, and V.~Koivunen, ``Generalized ambiguity function for
  the {MIMO} radar with correlated waveforms,'' in \emph{Proc. {IEEE} Int.
  Conf. Acoust., Speech,, Signal Process. (ICASSP)}, Florence, Italy, May 2014,
  pp. 5302--5306.

\bibitem{LevanonRadarSignal}
N.~Levanon and E.Mozeson, \emph{Radar Signals}.\hskip 1em plus 0.5em minus
  0.4em\relax Hoboken, NJ: Wiley, 2004.

\bibitem{SoltanalianTBDesign14}
M.~Soltanalian, H.~Hu, and P.~Stoica, ``Single-stage transmit beamforming
  design for {MIMO} radar,'' \emph{Signal Process.}, vol. 102, pp. 132--138,
  Mar. 2014.

\bibitem{RoweSpecConstraint14}
W.~Rowe, P.~Stoica, and J.~Li, ``Spectrally constrained waveform design,''
  \emph{{IEEE} Signal Process. Mag.}, vol.~31, no.~3, pp. 157--162, May 2014.

\bibitem{ForsytheMIMOWavCon10}
K.~W. Forsythe and D.~W. Bliss, ``{MIMO} radar waveform constraints for
  {GMTI},'' \emph{{IEEE} J. Sel. Topics Signal Process.}, vol.~4, no.~1, pp.
  21--32, Feb. 2010.

\bibitem{BlumWaveform07}
Y.~Yang and R.~S. Blum, ``{MIMO} radar waveform design based on mutual
  information and minimum mean-square error estimation,'' \emph{{IEEE} Trans.
  Aerosp. Electron. Syst.}, vol.~43, no.~1, pp. 330--343, Jan. 2007.

\bibitem{LiSignalSyn}
J.~Li, P.~Stoica, and X.~Zheng, ``Signal synthesis and receiver design for
  {MIMO} radar imaging,'' \emph{{IEEE} Trans. Signal Process.}, vol.~56, no.~8,
  pp. 3959--3968, Aug. 2008.

\bibitem{DengPolyphaseWaveform}
H.~Deng, ``Polyphase code design for orthogonal netted radar systems,''
  \emph{{IEEE} Trans. Signal Process.}, vol.~52, no.~11, pp. 3126--3135, Nov.
  2004.

\bibitem{RabideauWaveform08}
D.~J. Rabideau, ``Adaptive {MIMO} radar waveforms,'' in \emph{Proc. {IEEE}
  Radar Conf.}, Rome, Italy, May 2008, pp. 1--6.

\bibitem{SongWaveform10}
X.~Song, S.~Zhou, and P.~Willett, ``Reducing the waveform cross correlation of
  {MIMO} radar with space-time coding,'' \emph{{IEEE} Trans. Signal Process.},
  vol.~58, no.~8, pp. 4213--4224, Aug. 2010.

\bibitem{MaSidelobeSuppress}
C.~Ma, T.~S. Yeo, C.~S. Tan, Y.~Qiang, and T.~Zhang, ``Receiver design for
  {MIMO} radar range sidelobes suppression,'' \emph{{IEEE} Trans. Signal
  Process.}, vol.~58, no.~10, pp. 5469--5474, Oct. 2010.

\bibitem{HuaRangeDopplerSupp}
G.~Hua and S.~S. Abeysekera, ``Receiver design for range and sidelobe
  suppression using {MIMO} and phased-array radar,'' \emph{{IEEE} Trans. Signal
  Process.}, vol.~61, no.~6, pp. 1315--1326, Mar. 2013.

\bibitem{OptimumArrayPro}
H.~L. {Van Trees}, \emph{Optimum array processing}.\hskip 1em plus 0.5em minus
  0.4em\relax New York: Wiley, 2002.

\bibitem{XuBeamspaceESPRIT}
G.~Xu, S.~D. Silverstein, R.~H. Roy, and T.~Kailath, ``Beamspace {ESPRIT},''
  \emph{{IEEE} Trans. Signal Process.}, vol.~42, no.~2, pp. 349--356, Feb.
  1994.

\bibitem{CVX}
\BIBentryALTinterwordspacing
M.~Grant and S.~Boyd. (2014, Sep.) {CVX}: Matlab software for disciplined
  convex programming (version 2.1). [Online]. Available:
  \url{http://cvxr.com/cvx}
\BIBentrySTDinterwordspacing

\bibitem{VorobyovSAMTutorial}
\BIBentryALTinterwordspacing
S.~A. Vorobyov and M.~Lops. (2014, Jun.) Trade-offs in {MIMO} radar with
  colocated antennas. Tutorial in 8th {IEEE} Sensor Array and Multichannel
  Signal Process. Workshop. A Coru\~{n}a, Spain. [Online]. Available:
  \url{http://www.gtec.udc.es/sam2014/images/sam2014/tutorials/slides_tutorial_tradeoffs_mimo_radar_sam_2014_v2.pdf}
\BIBentrySTDinterwordspacing

\end{thebibliography}

\end{document}